\documentclass[journal]{IEEEtran}
\usepackage[T1]{fontenc}
\usepackage{ifpdf}
\usepackage{cite}
\usepackage[pdftex]{graphicx}
\usepackage{amsmath}
\usepackage{amssymb}
\usepackage{bm}
\usepackage{array}
\usepackage{fixltx2e}
\usepackage{stfloats}

\usepackage{balance} 

\usepackage{mathtools} 

\usepackage{diagbox}

\usepackage{makecell}

\usepackage{threeparttable}

\usepackage{xcolor,soul}
\sethlcolor{cyan}
\soulregister\cite7

\usepackage{algorithm}
\usepackage{algpseudocode}

\newtheorem{theorem}{Theorem}

\newtheorem{lemma}{Lemma}

\newtheorem{remark}{Remark}

\begin{document}
\title{\huge Joint IRS Location and Size Optimization in Multi-IRS Aided Two-Way Full-Duplex Communication Systems}

\author{Christos N. Efrem and Ioannis Krikidis, \IEEEmembership{Fellow, IEEE}
\thanks{This work was supported by the Research Promotion Foundation, Cyprus, under the project INFRASTRUCTURES/1216/0017 (IRIDA). This work was also supported by the European Research Council (ERC) under the European
Union's Horizon 2020 research and innovation programme (Grant agreement No. 819819). Part of this work was presented at the IEEE International Conference on Communications (ICC), 2022 \cite{Efrem_ICC}. 

The authors are with the Department of Electrical and Computer Engineering, University of Cyprus, 1678 Nicosia, Cyprus (e-mail: \{efrem.christos, krikidis\}@ucy.ac.cy). 

This article has been accepted for publication in \textit{IEEE Transactions on Wireless Communications}, February 2023. Copyright \copyright 2023 IEEE. Personal use is permitted, but republication/redistribution requires IEEE permission.  }}

\markboth{}%
{}

\maketitle

\begin{abstract}
Intelligent reflecting surfaces (IRSs) have emerged as a promising wireless technology for the dynamic configuration and control of electromagnetic waves, thus creating a smart (programmable) radio environment. In this context, we study a multi-IRS assisted two-way communication system consisting of two users that employ full-duplex (FD) technology. More specifically, we deal with the joint IRS location and size (i.e., the number of reflecting elements) optimization in order to minimize an upper bound of system outage probability under various constraints: minimum and maximum number of reflecting elements per IRS, maximum number of installed IRSs, maximum total number of reflecting elements (implicit bound on the signaling overhead) as well as maximum total IRS installation cost. First, the problem is formulated as a discrete optimization problem and, then, a theoretical proof of its NP-hardness is given. Moreover, we provide a lower bound on the optimum value by solving a linear-programming relaxation (LPR) problem. Subsequently, we design two polynomial-time algorithms, a deterministic greedy algorithm and a randomized approximation algorithm, based on the LPR solution. The former is a heuristic method that always computes a feasible solution for which (a posteriori) performance guarantee can be provided. The latter achieves an approximate solution, using randomized rounding, with provable (a priori) probabilistic guarantees on the performance. Furthermore, extensive numerical simulations demonstrate the superiority of the proposed algorithms compared to the baseline schemes. Finally, useful conclusions regarding the comparison between FD and conventional half-duplex (HD) systems are also drawn.

\end{abstract}


\IEEEpeerreviewmaketitle

\section{Introduction}

Intelligent reflecting surface (IRS), also known as reconfigurable intelligent surface, has been considered as one of the most effective techniques to cope with the signal blockage in communication networks operating in millimeter-wave (mmWave) frequency bands. IRS is a planar surface which is installed on the walls or ceilings of buildings so as to create virtual line-of-sight (LoS) links between the transmitters and receivers, thus overcoming the physical obstacles between them. In particular, IRS consists of (mostly) passive reflecting elements that can independently induce a controllable phase shift on the incident electromagnetic wave \cite{DiRenzo2019, Wu2020, Huang2020}. 

Moreover, IRSs are not expected to perform any sophisticated signal processing operations that require radio-frequency (RF) chains, but only the necessary amplitude attenuation and phase rotation of signals via low-power electronic circuits. In other words, high-cost active components (e.g., power amplifiers) are not required, thus leading to low energy consumption and implementation cost. For this reason, IRSs are usually referred to as ``nearly-passive'' devices. In addition, they have much lower implementation cost than conventional technologies of active transceivers, such as amplify-and-forward, decode-and-forward relays and multiple-input-multiple-output systems \cite{DiRenzo2020, Wu2021}. 

On the other hand, full-duplex (FD) wireless technology has the potential to double the spectral efficiency, compared to its half-duplex (HD) counterpart, by allowing simultaneous transmission and reception within the same frequency band. This can be achieved at the expense of higher implementation complexity due to the required loop-interference cancellation techniques \cite{Sabharwal2014, Zhang2015, Liu2015}. Recently, there is a growing interest of the research community in combining IRSs with FD systems in order to exploit their benefits and advantages \cite{Pan2021, Atapattu2020, Abdullah2021, Peng2021, Saeidi2021, Zhang2020, Xu2020}.

\subsection{Related Work}

To begin with, \cite{Lu2021} presents an aerial-IRS (AIRS) system architecture in order to enhance the system performance, compared to the conventional terrestrial IRS, by exploiting the high altitude of AIRS. In particular, the authors study the joint optimization of the transmit beamforming of the ground source-node as well as the placement and passive beamforming of the AIRS. In addition, the single-IRS deployment problem (inside a 3-dimensional box), where an access point communicates with multiple users via the IRS, has been investigated in \cite{Mu2021}. Specifically, the weighted sum rate maximization problem has been formulated for three multiple access schemes: non-orthogonal multiple access (NOMA), frequency division multiple access (FDMA), and time division multiple access (TDMA). In order to deal with these problems, the authors have used several methods, namely, monotonic optimization, semidefinite relaxation, alternating optimization and successive convex approximation. The capacity region of an IRS-aided communication system with two users has been studied in \cite{S_Zhang2021}, for centralized and distributed (in two locations) IRS deployments. In millimeter-wave networks assisted by a single IRS, the optimal system
performance is achieved when the IRS is placed closer to the receiver than the transmitter \cite{Ntontin2021}.
 
Furthermore, the problem of joint IRS deployment, phase-shift design as well as power allocation for maximizing the energy efficiency of a NOMA network has been recently formulated and solved using machine learning methods \cite{Liu2021}. As concerns the coverage of an IRS-assisted network with one base station and one user equipment, \cite{Zeng2021} has examined the IRS placement problem to maximize the cell coverage by optimizing the IRS orientation and horizontal distance from the base station. Also, the optimal number of reflecting elements for an IRS, assisting the communication between a transmitter and a receiver, has been proposed in \cite{Zappone2021}. In particular, the system rate, energy efficiency, and their tradeoff are maximized by taking into consideration the signaling overhead required for the channel estimation and IRS phase-shift configuration. 

Moreover, the (single) FD relay location and power optimization problem in a point-to-point (P2P) communication system has already been addressed in \cite{Yu2015, Li2017, Hou2018}. Finally, \cite{Lyras2018, Lyras2020, Efrem2020, Efrem2021} have developed efficient optimization algorithms for the deployment of ground stations in RF and optical satellite networks with site diversity.

\subsection{Main Contributions}

In the majority of existing works, the IRS positions are assumed to be known \emph{in advance}. However, IRS locations have a great impact on the overall system performance. As a result, their optimization is extremely important and deserves its own study. In this paper, we design efficient (polynomial-time) algorithms to jointly optimize the location and size (i.e., the number of reflecting elements) of multiple distributed IRSs in a two-way FD communication network. Specifically, the major contributions of this work are the following: 

\begin{itemize}
\item Extension of the IRS system model introduced in \cite{Atapattu2020} to multi-IRS systems, including not only small-scale fading but also large-scale path loss. In this way, we exploit the \emph{geometric characteristics} (i.e., the distances between users and IRSs) of the wireless network. Specifically, the deployment of multiple IRSs has two attractive features: i) \emph{small-scale diversity} between the reflecting elements of the same IRS, and ii) \emph{large-scale diversity} between the reflecting elements of distinct IRSs. In addition, the channel coefficients of a given user-to-IRS or IRS-to-user link can follow an \emph{arbitrary} probability distribution (not necessarily Rayleigh fading as in \cite{Atapattu2020}), while distinct IRSs may have different channel distributions. 

\item Recent works dealing with the IRS deployment often assume a \emph{continuous (bounded/unbounded) area for installing an IRS} (for example, \cite{Mu2021}, \cite{Ntontin2021}, and \cite{Zeng2021}). Unlike previous research, in this article we consider a \emph{predetermined and finite set of available IRS locations}, thus taking into account physical constraints for the IRS positions. This is of great practical interest, since IRSs are usually installed on the facades, walls or ceilings of existing buildings. Nevertheless, if we are interested in installing IRSs within a bounded continuous area, then this region can be divided into a sufficiently large finite number of distinct points (this method is known as discretization). Therefore, the proposed methodology is still applicable. 

\item Mathematical formulation of a discrete optimization problem in order to minimize an upper bound of system outage probability, which is subject to several constraints: minimum and maximum number of reflecting elements for each IRS, maximum number of installed IRSs, maximum total number of reflecting elements and maximum total IRS installation cost. In addition, a theoretical proof of its computational complexity (\emph{NP-hardness}) is given.

\item Furthermore, we construct a \emph{linear-programming relaxation (LPR)} so as to lower bound the optimum value. Then, we develop two polynomial-time algorithms, namely, a \emph{deterministic greedy algorithm} and a \emph{randomized approximation algorithm}, whose key ingredient is the LPR solution. The first is a heuristic method which always finds a feasible solution to the problem, while the second achieves an approximate solution via randomized rounding. For the randomized algorithm, we also provide \emph{probabilistic performance guarantees} using concentration inequalities (in particular, Hoeffding's bound). 

\item Finally, numerical results show the superiority of the proposed algorithms compared to the benchmarks, while \emph{useful comparisons between FD and HD schemes} are provided as well.
 
\end{itemize}

\subsection{Outline and Notation}

The remainder of this paper is organized as follows. Section \ref{section:System_model} describes the system model, while Section \ref{section:Problem_formulation} formulates the optimization problem and studies its computational complexity. Afterwards, Section \ref{section:Optimization_algorithms} develops and analyzes the proposed optimization algorithms. In addition, numerical results are provided in Section \ref{section:Numerical_results}. Finally, useful conclusions and future research directions are given in Section \ref{section:Conclusion}, while Appendices \ref{appendix:NP_hardness}, \ref{appendix:Expectation_guarantees} and \ref{appendix:Deviation_guarantees} contain the proofs of theorems.

\textit{Mathematical notation}: Italic letters denote (real/complex) scalars, boldface letters represent vectors and matrices, while calligraphic letters stand for sets and events. $\left| z \right|$ denotes the absolute value (or magnitude) of a complex number $z$ and $j = \sqrt{- 1}$ is the imaginary unit. In addition, $\left| \mathcal{A} \right|$ and $\mathcal{B}^{\mathsf{c}}$ represent the cardinality of a set $\mathcal{A}$ and the complement of an event $\mathcal{B}$, respectively. The Cartesian product of the sets ${\left\{ {{\mathcal{A}_n}} \right\}_{n \in \mathcal{N}}} = \{ {\mathcal{A}_1} , \dots , {\mathcal{A}_N} \} $ is denoted by $\times_{n \in \mathcal{N}}{\mathcal{A}_n} = {\mathcal{A}_1} \times \cdots \times {\mathcal{A}_N}$. Moreover, ${{\mathbf{0}}_N}$ is the $N$-dimensional zero vector and $\left[\cdot \right]^\top$ stands for the matrix transpose. The symbols $ \triangleq $ and $ \sim $ mean ``equal by definition'' and ``distributed as'', respectively. Also, $\log( \cdot )$ represents the natural logarithm (i.e., with base $e$) and $\binom{n}{m} = \tfrac{n!}{m!(n - m)!}$ is the binomial coefficient. $\Theta(\cdot)$, $O(\cdot)$, $\Omega(\cdot)$ and $o(\cdot)$ are respectively the big-theta, big-oh, big-omega and little-oh asymptotic notation. Furthermore, the floor and ceiling functions are denoted by $\lfloor \cdot \rfloor$ and $\lceil \cdot \rceil$, respectively. For every $x \geq 0$, $\operatorname{frac}(x) = x - \lfloor x \rfloor$ is the fractional part of $x$, with $\operatorname{frac}(x) \in [0,1)$, and $\operatorname{round}(x) = \lfloor {x + 0.5} \rfloor$. In addition, $\Pr \left( \cdot \right)$ and $\mathbb{E}\left( \cdot \right)$ denote probability and expectation (or expected value), respectively. Finally, $\operatorname{Uniform}\left( \mathcal{D} \right)$ stands for the continuous/discrete uniform distribution on the set $\mathcal{D}$.

\section{System Model} \label{section:System_model}

\begin{figure}[!t]
\centering
\includegraphics[width=3.5in]{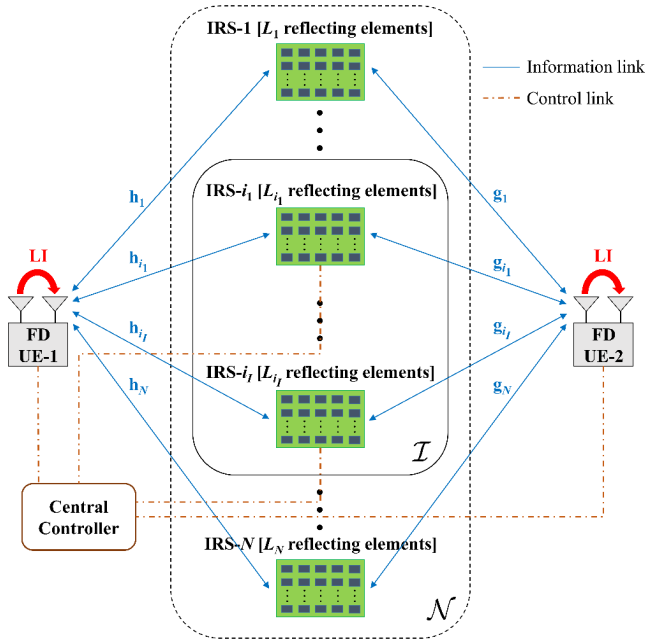} 
\caption{Multi-IRS system assisting two-way FD communication with reciprocal channels and negligible direct link. The set of available IRS locations, $\mathcal{N}$, and the set of finally installed IRSs, $\mathcal{I}$, are illustrated by the dashed-outline and solid-outline rectangles, respectively. In each time-slot, the central controller activates only one IRS form the set $\mathcal{I}$, while the remaining IRSs are idle (i.e., non-reflective).}
\label{fig:system_config}
\end{figure}

In this paper, we deal with a multi-IRS system assisting a two-way P2P communication link, as shown in Fig. \ref{fig:system_config}, where the locations of users are \emph{fixed} (i.e., the two users do not move). In particular, each user equipment (UE) operates in FD mode and therefore is equipped with either a single shared-antenna or a pair of separate antennas for signal transmission and reception, depending on the FD implementation \cite{Sabharwal2014}. In addition, $\mathcal{N} = \{ 1, \ldots ,N\} $ represents the set of available locations for installing an IRS (with $N \geq 1$), while $\mathcal{I} = \{ {i_1}, \ldots ,{i_I}\}  \subseteq \mathcal{N}$ (where $I = \left| \mathcal{I} \right|$) stands for the set of finally installed IRSs. In each time-slot, we assume that exactly one IRS from the set $\mathcal{I}$ is active and the remaining IRSs are idle (i.e., non-reflective). UE-1 transmits its data to UE-2, through the active IRS, and UE-2 transmits its data to UE-1, through the same IRS, simultaneously (i.e., within the same time-slot) by using the same frequency band. Note that FD technology is \emph{actually} applied at both UEs, whereas IRSs are treated as passive devices that \emph{inherently} operate in FD \cite{Atapattu2020}. The transmit power of each UE is considered \emph{fixed} for all time-slots; power control is outside the scope of this paper. Since both UEs suffer from strong loop-interference (LI) due to the FD operation, they employ the same LI-cancellation techniques (e.g., passive and active suppression in the analog/digital domain \cite{Sabharwal2014, Zhang2015, Liu2015}), resulting in residual LI.   

Moreover, the total UE-to-IRS and IRS-to-UE transmission time is within a coherence interval of the wireless channel. As a result, the forward and backward channels between a UE and an IRS can be regarded almost identical (\emph{reciprocal channels}) \cite{Atapattu2020}. Also, the direct link between UEs is considered strongly attenuated (high path-loss) due to the long distance, high carrier-frequency, or severe blockage by physical obstacles (\emph{no direct link}). All IRSs are assumed to be passive (i.e., performing only phase shifts without amplification) and they have negligible delay regarding the reflection of incident electromagnetic waves. We assume \emph{perfect channel state information (CSI)}, i.e., without estimation errors, and \emph{global CSI knowledge}, i.e., available to both UEs \cite{Atapattu2020}.\footnote{IRS-assisted systems with imperfect CSI have been studied in \cite{S_Hu2021}. The extension of our methodology to such a case deserves further investigation.} Furthermore, there is a \emph{central controller} that performs the IRS activation, adjusts the IRS phase-shifts and communicates the necessary CSI knowledge between UEs via separate low-latency wireless/wired backhaul links (illustrated by dash-dotted lines in Fig. \ref{fig:system_config}).

Let ${L_n}$ be the number of reflecting elements of the ${n^{{\text{th}}}}$ IRS. The channel coefficient from UE-1 (UE-2) to the ${\ell ^{{\text{th}}}}$ reflecting element of the ${n^{{\text{th}}}}$ IRS is denoted by ${h_{n,\ell }} = \left| {{h_{n,\ell }}} \right|{e^{j{\vartheta _{n,\ell }}}}$ (respectively, ${g_{n,\ell }} = \left| {{g_{n,\ell }}} \right|{e^{j{\psi _{n,\ell }}}}$), for every $n \in \mathcal{N}$ and $\ell  \in {\mathcal{L}_n} = \{ 1, \ldots ,{L_n}\}$. Also, the channel coefficients remain constant during one time-slot, but they change independently between distinct time-slots. For notational convenience, the channel coefficients corresponding to the ${n^{{\text{th}}}}$ IRS can be grouped in vector form, i.e., ${{\mathbf{h}}_n} = {[{h_{n,1}}, \ldots ,{h_{n,{L_n}}}]^ \top }$ and ${{\mathbf{g}}_n} = {[{g_{n,1}}, \ldots ,{g_{n,{L_n}}}]^ \top }$. All channel coefficients are assumed to be \emph{mutually independent} \cite{Atapattu2020}, while, for a given $n \in \mathcal{N}$, $\mathbf{h}_n$ and $\mathbf{g}_n$ are individually \emph{independent and identically distributed (i.i.d.)} with possibly different probability distributions (e.g., Rice and Rayleigh distributions, respectively). The diagonal (${L_n} \times {L_n}$) phase-shift matrix of the ${n^{{\text{th}}}}$ IRS is given by ${{\mathbf{\Phi }}_n} = {\text{diag}}({e^{j{\phi _{n,1}}}}, \ldots ,{e^{j{\phi _{n,{L_n}}}}})$, i.e., we consider only phase-shifts and not amplitude attenuation.\footnote{There are practical models where the amplitude and phase-shift of IRS elements are dependent on each other, e.g., \cite{X_Yu2021}. However, these models are beyond the scope of this paper.}

Under the above assumptions and following a similar approach with \cite{Atapattu2020}, the received signals at UE-1 and UE-2 in time-slot $t$ (after the LI mitigation), when only the $n^\text{th}$ IRS is active, are expressed as follows 
\begin{equation}
\begin{split}
  {y_1}(t) = & \sqrt {{P_2}} \sqrt {{\delta _{n,2}}{\delta _{n,1}}} {\mathbf{g}}_n^ \top {{\mathbf{\Phi }}_n}{{\mathbf{h}}_n}{s_2}(t)  \\
  & + \sqrt {{P_1}} {\delta _{n,1}}{\mathbf{h}}_n^ \top {{\mathbf{\Phi }}_n}{{\mathbf{h}}_n}{s_1}(t) + {\xi _1}(t) + {w_1}(t) ,
\end{split} 
\end{equation}
\begin{equation}
\begin{split}
  {y_2}(t) = & \sqrt {{P_1}} \sqrt {{\delta _{n,1}}{\delta _{n,2}}} {\mathbf{h}}_n^ \top {{\mathbf{\Phi }}_n}{{\mathbf{g}}_n}{s_1}(t) \\
  & + \sqrt {{P_2}} {\delta _{n,2}}{\mathbf{g}}_n^ \top {{\mathbf{\Phi }}_n}{{\mathbf{g}}_n}{s_2}(t) + {\xi _2}(t) + {w_2}(t)  ,
\end{split}
\end{equation}
where ${P_k}$, ${s_k}(t)$, ${\xi _k}(t)$ and ${w_k}(t)$ are the transmit power, information symbol, residual LI and additive white Gaussian noise (AWGN) of UE-$k$, respectively, for $k \in \{ 1,2\} $. In addition, ${\delta _{n,k}} = {A_0}d_{n,k}^{ - \alpha }$ accounts for the large-scale path loss between the ${n^{{\text{th}}}}$ IRS and UE-$k$, where ${A_0}$ is a positive constant that depends on the carrier frequency, ${d_{n,k}}$ is their Euclidean distance, and $\alpha $ is the path-loss exponent which depends on the wireless propagation environment. Note that, in the above equations, the first term represents the desired signal, while the second term is the self-interference (SI) induced by the IRS reflection of users' own transmitted symbols. Given that UE-$k$ has knowledge of ${P_k}$, ${s_k}(t)$, ${\delta _{n,k}}$, ${{\mathbf{h}}_n}$ (required for $k = 1$), ${{\mathbf{g}}_n}$ (needed for $k = 2$), and ${{\mathbf{\Phi }}_n}$, it can completely remove the SI. Moreover, the residual LI ${\xi _k}(t)$ and AWGN ${w_k}(t)$ are modeled as independent zero-mean complex Gaussian random variables with variances $\sigma _{{\text{L}}{{\text{I}}_k}}^2$ and $\sigma _{{w_k}}^2$, respectively. The variance of ${\xi _k}(t)$ can be further expressed as $\sigma _{{\text{L}}{{\text{I}}_k}}^2 = \omega P_k^\nu $, where the constants $\omega  > 0$ and $\nu  \in [0,1]$ depend on the LI cancellation technique applied at the UEs \cite{Atapattu2020}.  

For the sake of simplicity, we assume the following: ${P_1} = {P_2} = P$ (the same transmit power),  $\mathbb{E}( {{{\left| {{s_1}(t)} \right|}^2}} ) = \mathbb{E} ( {{{\left| {{s_2}(t)} \right|}^2}} ) = 1$ (unit-power information symbols), $\sigma _{{w_1}}^2 = \sigma _{{w_2}}^2 = \sigma _w^2$ and $\sigma _{{\text{L}}{{\text{I}}_1}}^2 = \sigma _{{\text{L}}{{\text{I}}_2}}^2 = \sigma _{{\text{LI}}}^2( = \omega {P^\nu })$ (equal noise and residual-LI power). Consequently, the \emph{instantaneous} signal-to-interference-plus-noise ratio (SINR) at both UEs, after the SI elimination, when communicating via the ${n^{{\text{th}}}}$ IRS is given by\footnote{In order to derive the SINR formula, observe that ${\mathbf{g}}_n^ \top {{\mathbf{\Phi }}_n}{{\mathbf{h}}_n} = {({\mathbf{g}}_n^ \top {{\mathbf{\Phi }}_n}{{\mathbf{h}}_n})^ \top } = {\mathbf{h}}_n^ \top {{\mathbf{\Phi }}_n}{{\mathbf{g}}_n} = \sum_{\ell  \in {\mathcal{L}_n}} {{h_{n,\ell }}{e^{j{\phi _{n,\ell }}}}{g_{n,\ell }}}  = \sum_{\ell  \in {\mathcal{L}_n}} {\left| {{h_{n,\ell }}} \right|\left| {{g_{n,\ell }}} \right|{e^{j({\phi _{n,\ell }} + {\vartheta _{n,\ell }} + {\psi _{n,\ell }})}}}$.}
\begin{equation}
{\gamma _n} = {\rho _n}{\left| {\sum\limits_{\ell  \in {\mathcal{L}_n}} {\left| {{h_{n,\ell }}} \right|\left| {{g_{n,\ell }}} \right|{e^{j({\phi _{n,\ell }} + {\vartheta _{n,\ell }} + {\psi _{n,\ell }})}}} } \right|^2} ,
\end{equation}
where
\begin{equation} \label{equation:rho_n}
{\rho _n} = \frac{{P{\delta _n}}}{{\sigma _{{\text{LI}}}^2 + \sigma _w^2}},
\end{equation}
with ${\delta _n} = {\delta _{n,1}}{\delta _{n,2}}$ being the overall path-loss between the two UEs through the ${n^{{\text{th}}}}$ IRS. This SINR formula is quite similar to that in \cite{Atapattu2020}, except for the total path-loss term ${\delta _n}$ that is explicitly included in $\rho_n$ instead of being incorporated in the channel coefficients $h_{n,\ell}$ and $g_{n,\ell}$.\footnote{Herein, however, we study a generalization of the system configuration presented in \cite{Atapattu2020}, including multiple IRSs and exploiting their geometric characteristics. In addition, the channel coefficients can follow \emph{any} probability distribution, not necessarily Rayleigh fading.}  

Furthermore, the IRS phase-shifts are optimally designed in order to maximize the instantaneous SINR, that is, 
\begin{equation} \label{equation:IRS_phase-shifts}
\phi _{n,\ell }^ \star  =  - {\vartheta _{n,\ell }} - {\psi _{n,\ell }},\;\; \forall \ell  \in {\mathcal{L}_n} = \{ 1, \ldots ,{L_n}\} .
\end{equation}
Note that the IRS phase-shifts are adjusted by the central controller after obtaining the necessary CSI knowledge (channel coefficients' phases) from the UE that performs the channel estimation. Also, we assume IRS phase-shifts \emph{without quantization errors}, i.e., the IRS phase-shift resolution is infinite; in practice, if the number of bits, $B$, used for controlling the phase of a reflecting element is very large (resulting in $2^B$ possible discrete values), then the quantization error can be considered insignificant. 

Therefore, the maximum SINR at both UEs (when communicating via the $n^\text{th}$ IRS) is written as follows 
\begin{equation} \label{equation:max_SINR}
\gamma _n^ \star  = {\rho _n}{\left( {\sum\limits_{\ell  \in {\mathcal{L}_n}} {\left| {{h_{n,\ell }}} \right|\left| {{g_{n,\ell }}} \right|} } \right)^2} = {\rho _n}{\zeta _n^2} \, ,
\end{equation}
where ${\zeta _n} = \sum_{\ell  \in {\mathcal{L}_n}} {{\zeta _{n,\ell }}} $, with ${\zeta _{n,\ell }} = \left| {{h_{n,\ell }}} \right|\left| {{g_{n,\ell }}} \right| \geq 0$, $\forall \ell  \in {\mathcal{L}_n}$. Observe that the random variables $\{ {\zeta _{n,\ell }} \}_{\ell \in \mathcal{L}_n}$ are \emph{i.i.d.}, because $\mathbf{h}_n$ and $\mathbf{g}_n$ are mutually independent and individually i.i.d.. Next, we denote the cumulative distribution function (CDF) of each $\zeta _{n,\ell}$ by $F_n: [0,+\infty) \rightarrow [0,1]$, i.e.,\footnote{If it is difficult to obtain an exact formula of $F_n(\cdot)$, we can use the \emph{moment matching} technique to derive an approximation by equating the mean and variance of the two distributions. For instance, the Gamma distribution can approximate several complicated distributions by tuning its shape and scale parameters \cite{Atapattu2020}.}     
\begin{equation} \label{equation:CDF_single_element}
F_n(u) \triangleq \Pr ( {\zeta _{n,\ell }} \leq u ),\;\;  \forall \ell \in \mathcal{L}_n .
\end{equation}

\vspace{2mm}
\begin{remark}[Rayleigh fading] \label{remark:Rayleigh_fading}
If $\mathbf{h}_n$ and $\mathbf{g}_n$ are i.i.d. complex normal/Gaussian random variables with zero mean and variance ${\sigma ^2}$, then $\left| {{h_{n,\ell }}} \right|,\left| {{g_{n,\ell }}} \right|\sim{\text{Rayleigh}}(\sigma / \sqrt 2)$. In addition, according to \cite{Atapattu2020}, the CDF of each Rayleigh-product random variable $\zeta _{n,\ell}$ is given by       
\begin{equation} \label{equation:CDF_single_element_Rayleigh}
F_n^{\text{R}}(u) = 1 - \frac{{2u}}{{{\sigma ^2}}}{K_1}\left( {\frac{{2u}}{{{\sigma ^2}}}} \right) ,
\end{equation}
where ${K_1 }( \cdot )$ is the modified Bessel function of the second kind of the first order.       
\end{remark}
\vspace{2mm}

Given an SINR threshold ${\gamma _{{\text{th}}}}$, the outage probability of each UE is defined as \linebreak ${P_{{\text{out}},n}}({L_n}) \triangleq \Pr ( \gamma _n^ \star  \leq {\gamma _{{\text{th}}}} )  = \Pr ( {\zeta _n} \leq \sqrt {{\gamma _{{\text{th}}}}/{\rho _n}} ) $. Unfortunately, an exact closed-form expression of ${P_{{\text{out}},n}}({L_n})$ is usually not available in general (or hard to compute even if it exists). Nevertheless, to circumvent this difficulty, we resort to an upper bound of outage probability, i.e., 
\begin{equation} \label{equation:upper_bound_outage_probability}
\begin{split}
{P_{{\text{out}},n}}({L_n}) & = \Pr \left( {\sum\limits_{\ell  \in {\mathcal{L}_n}} {{\zeta _{n,\ell }}}  \leq \sqrt {{\gamma _{{\text{th}}}}/{\rho _n}} } \right) \\
& \mathop  \leq \limits^{({\text{a}})} \Pr \left( {\mathop {\max }\limits_{\ell  \in {\mathcal{L}_n}} \{ {\zeta _{n,\ell }}\}  \leq \sqrt {{\gamma _{{\text{th}}}}/{\rho _n}} } \right)  \\
& = \Pr \left( {\bigcap\limits_{\ell  \in {\mathcal{L}_n}} {\{ {\zeta _{n,\ell }} \leq \sqrt {{\gamma _{{\text{th}}}}/{\rho _n}} \} } } \right)  \\
& \mathop  = \limits^{({\text{b}})}   \prod\limits_{\ell  \in {\mathcal{L}_n}} {\Pr ( {\zeta _{n,\ell }} \leq \sqrt {{\gamma _{{\text{th}}}}/{\rho _n}} ) }     \\
& \mathop  = \limits^{({\text{c}})}  {\left[ {F_n (\sqrt {{\gamma _{{\text{th}}}}/{\rho _n}} )} \right]^{{L_n}}} \triangleq {{\overline P}_{{\text{out}},n}}({L_n})  , 
\end{split}
\end{equation}
where inequality (a) is due to the fact that $\sum\nolimits_{\ell  \in {\mathcal{L}_n}} {{\zeta _{n,\ell }}}  \geq {\max _{\ell  \in {\mathcal{L}_n}}}\{ {\zeta _{n,\ell }}\} $, while equalities (b) and (c) follows from the independence of random variables ${\{ {\zeta _{n,\ell }}\} _{\ell  \in {\mathcal{L}_n}}}$ and equation \eqref{equation:CDF_single_element}, respectively. It is interesting to observe that the upper bound holds for \emph{any} (arbitrary) CDF $F_n(\cdot)$, while ${P_{{\text{out}},n}}({L_n})$ and ${\overline P_{{\text{out}},n}}({L_n})$ are both \emph{nonincreasing functions of} ${L_n}$. In addition, ${\lim _{{L_n} \to \infty }}{P_{{\text{out}},n}}({L_n}) = {\lim _{{L_n} \to \infty }}{\overline P_{{\text{out}},n}}({L_n}) = 0$, provided that $F_n(\sqrt {{\gamma_{{\text{th}}}}/{\rho_n}} ) < 1$.

\subsection{IRS Activation Policy}

As we mentioned earlier, exactly one IRS is activated in each time-slot by the central controller, while the remaining IRSs are inactive (i.e., non-reflective).\footnote{The activation of multiple IRSs within the same time-slot would cause inter-IRS/secondary reflections of the transmitted signal (i.e., between the active IRSs) as well as additional IRS-to-user SI links due to the FD operation. If IRSs are located close to each other, then secondary reflections are significant and might degrade the system performance due to increased interference. Moreover, SI elimination at each user would require extra signaling overhead and processing delay. As a result, the activation of a single IRS per time-slot does not impose any restrictions on the distances between IRSs (to ensure negligible secondary reflections), and also maintains the communication overhead and delay for SI cancellation as low as possible. Finally, synchronization issues related to various path-loss distances are more critical in multi-IRS activation scenarios.} In particular, the central controller activates the IRS that achieves the highest (instantaneous) SINR among the installed IRSs \cite{Mensi2022}, i.e.,  
\begin{equation} \label{equation:IRS_activation}
{i^ \star } \in \mathop {\arg \max }\limits_{i \in \mathcal{I}} \{ \gamma _i^ \star \} \;\; \Leftrightarrow \;\;\gamma _{{i^ \star }}^ \star  = \mathop {\max }\limits_{i \in \mathcal{I}} \{ \gamma _i^ \star \} ,
\end{equation}
where $\gamma_i^ \star$ is given by \eqref{equation:max_SINR}.

\subsection{Upper Bound of System Outage Probability}

Based on the aforementioned IRS activation strategy, the \emph{system outage probability} can be computed as follows 
\begin{equation} \label{equation:system_outage_probability}
\begin{split}
{P_{{\text{out}}}}(\mathcal{I},{\mathbf{L}}) & \triangleq \Pr ( \gamma _{{i^ \star }}^ \star  \leq {\gamma _{{\text{th}}}}) \mathop  = \limits^{({\text{d}})} \Pr \left( {\mathop {\max }\limits_{i \in \mathcal{I}} \{ \gamma _i^ \star \}  \leq {\gamma _{{\text{th}}}}} \right)  \\
&  = \Pr \left( {\bigcap\limits_{i \in \mathcal{I}} {\{ \gamma _i^ \star  \leq {\gamma _{{\text{th}}}}\} } } \right) \mathop  = \limits^{({\text{e}})} \prod\limits_{i \in \mathcal{I}} {\Pr ( \gamma _i^ \star  \leq {\gamma _{{\text{th}}}} ) }   \\
& \mathop  = \limits^{({\text{f}})}  \prod\limits_{i \in \mathcal{I}} {{P_{{\text{out}},i}}({L_i})} ,
\end{split}
\end{equation}
where ${\mathbf{L}} = {[{L_1}, \ldots ,{L_N}]^ \top }$ and ${\gamma _{{\text{th}}}}$ is the SINR threshold. Equalities (d), (e) and (f) follow from \eqref{equation:IRS_activation}, the independence of  ${\{ \gamma _i^ \star \} _{i \in \mathcal{I}}}$ (due to the independence of ${\{ {\zeta _i}\} _{i \in \mathcal{I}}}$) and the definition of ${P_{{\text{out}},i}}({L_i}) \triangleq \Pr ( \gamma _i^ \star  \leq {\gamma _{{\text{th}}}} )$, respectively.

Afterwards, by combining \eqref{equation:system_outage_probability} with \eqref{equation:upper_bound_outage_probability}, we obtain the following upper bound of system outage probability\footnote{In optimization theory, it is a standard approach to derive and minimize an upper bound (in case of minimization problems) when the original objective function is hard to compute (e.g., in closed form). In addition, the proposed bound is suitable for constructing a \emph{relaxation problem that is solvable in polynomial time} (see Section \ref{subsection:LPR}), thus achieving mathematical tractability. At this point, we would like to emphasize that the tightness of the upper bound (which depends on the probability distribution of channel coefficients) is \emph{not} the main concern here. Instead, from the optimization perspective, the upper bound should satisfy a \emph{weaker condition}: if $(\mathcal{I}^\star,{\mathbf{L}}^\star) \in \arg \min_{(\mathcal{I},{\mathbf{L}}) \in \mathcal{C}} {\overline P_{{\text{out}}}}(\mathcal{I},{\mathbf{L}})$, then ${P_{{\text{out}}}}(\mathcal{I}^\star,{\mathbf{L}}^\star) \approx \min_{(\mathcal{I},{\mathbf{L}}) \in \mathcal{C}} {P_{{\text{out}}}}(\mathcal{I},{\mathbf{L}})$, where $\mathcal{C}$ is the constraint set. In other words, any solution that minimizes the upper bound should also (approximately) minimize the exact outage probability, without necessarily requiring ${\overline P_{{\text{out}}}}(\mathcal{I}^\star,{\mathbf{L}}^\star) \approx {P_{{\text{out}}}}(\mathcal{I}^\star,{\mathbf{L}}^\star)$. Although it is quite difficult to check such a condition, the following example provides some evidence about the suitability of the upper bound. Given any fixed $\mathcal{I} \subseteq \mathcal{N}$, it holds that   ${\mathbf{L}^{\max}} \in \{ \arg \min_{{\mathbf{L}} \in \mathcal{C}_\mathbf{L}} {\overline P_{{\text{out}}}}(\mathcal{I},{\mathbf{L}}) \} \cap \{ \arg \min_{{\mathbf{L}} \in \mathcal{C}_\mathbf{L}} {P_{{\text{out}}}}(\mathcal{I},{\mathbf{L}}) \}$, where ${\mathbf{L}^{\max}} = {[{L_1^{\max}}, \ldots ,{L_N^{\max}}]^\top}$ and $\mathcal{C}_\mathbf{L} = \{ \mathbf{L} : \, {L_n} \in \{ L_n^{\min }, \ldots ,L_n^{\max }\}, \forall n \in \mathcal{N} \}$, since ${\overline P_{{\text{out}}}}(\mathcal{I},{\mathbf{L}})$ and ${P_{{\text{out}}}}(\mathcal{I},{\mathbf{L}})$ are nonincreasing in each $L_n$. The physical interpretation is that, under only individual size constraints, the minimization of exact outage probability (and its upper bound) is achieved when the number of reflecting elements is the maximum possible.} 
\begin{equation} \label{equation:upper_bound_system_outage_probability}
{P_{{\text{out}}}}(\mathcal{I},{\mathbf{L}}) \leq \prod\limits_{i \in \mathcal{I}} {{{\left[ {F_i ( \sqrt {{\gamma _{{\text{th}}}}/{\rho _i}} )} \right]}^{{L_i}}}}  \triangleq {\overline P_{{\text{out}}}}(\mathcal{I},{\mathbf{L}}) .
\end{equation}

\subsection{IRS Installation Cost Model}

In this paper, we model the installation cost of IRS $n \in \mathcal{N}$ as an \emph{affine function} of the number of reflecting elements, i.e.,
\begin{equation} \label{equation:single_IRS_cost}
{C_n}({L_n}) = {c_n} + {\lambda _n}{L_n} ,
\end{equation}
where ${c_n} \geq 0$ is the fixed deployment cost and ${\lambda _n} \geq 0$ is the cost rate (measured in cost-units per element) of the corresponding IRS.\footnote{Note that ${c_n}$ includes the rent of IRS location for a specific period of time, while ${\lambda _n}$ can be obtained from IRS manufacturers and is expected to be the same for all IRSs (in general, it may be different for distinct IRSs).} In addition, the total installation cost is defined as the sum of the costs of all IRSs in the set $\mathcal{I}$, i.e., 
\begin{equation} \label{equation:total_IRS_cost}
{C_{{\text{tot}}}}(\mathcal{I},{\mathbf{L}}) = \sum\limits_{i \in \mathcal{I}} {{C_i}({L_i})} = \sum\limits_{i \in \mathcal{I}} {({c_i} + {\lambda _i}{L_i})} .
\end{equation}

\vspace{2mm}
\begin{remark}
In general, the IRS installation cost can be any nondecreasing function of the number of reflecting elements. In this case, we can approximate the installation cost by an affine function given by \eqref{equation:single_IRS_cost}. In particular, the coefficients $\{{c_n},{\lambda _n}\}$ can be adjusted so as to minimize the error between the two functions (a procedure known as \emph{curve fitting}) within a given interval of interest, e.g., for \linebreak ${L_n} \in \{ L_n^{\min }, \ldots ,L_n^{\max }\}$.
\end{remark}

\subsection{Implementation Issues} \label{subsection:Implementation}

First of all, the central controller is responsible for the synchronization of the following operations: 1) channel estimation for all the installed IRSs, 2) transfer of the total CSI knowledge from UE-2, that performs the channel estimation, to the central controller, 3) IRS activation and phase-shift adjustments (performed by the central controller, according to equations \eqref{equation:IRS_phase-shifts}, \eqref{equation:max_SINR} and \eqref{equation:IRS_activation}) plus notification of these decisions to both UEs, and 4) transfer of CSI corresponding to the activated IRS from the central controller to UE-1. It is very important for UEs to know which IRS is active, in each time-slot, together with the corresponding channel coefficients and phase-shift matrix in order to achieve complete cancellation of the SI. Consequently, the overall required overhead is $\Theta(L_{\text{tot}})$, where $L_{\text{tot}}$ is the total number of reflecting elements. 

Regarding the channel estimation of the (installed) IRSs, we can utilize $I = \left| \mathcal{I} \right|$ distinct orthogonal resource-blocks (e.g., frequency bands or time-slots), one for each IRS.  Suppose that we want to estimate the channels between the IRS $i \in \mathcal{I}$ and UEs. In the corresponding resource-block, all the remaining IRSs are kept inactive (i.e., non-reflective), UE-1 acts only as a transmitter and UE-2 acts only as a receiver. In particular, UE-2 estimates the pair of channel coefficients $\left\{ {\{ {\delta _{i,1}},{{\mathbf{h}}_i}\} ,\{ {\delta _{i,2}},{{\mathbf{g}}_i}\} } \right\}$ based on the received signal (observation), assuming that the transmitted symbols of UE-1 and IRS phase-shifts (pilot signals) are known to the receiver (UE-2); more details about the channel estimation procedure can be found in \cite{He2020}.

\subsection{Half-Duplex (HD) Scheme} \label{subsection:HD}

If the system operates in half-duplex (HD) mode, then the UEs transmit their data in two distinct time-slots: time-slot 1 is allocated for UE-1 transmission and time-slot 2 for UE-2 transmission. As a result, each UE is equipped with a single antenna, utilizes the same frequency band, and there is no interference at all (neither self nor loop interference). 

In this case, we can study the performance of HD scheme by appropriately modifying the previous equations of FD scheme. In particular, we should replace ${\rho _n}$ in \eqref{equation:rho_n} with $\rho _n^{{\text{HD}}} = {{P{\delta _n}} / {\sigma _w^2}}$ (since $\sigma _{{\text{LI}}}^2 = 0$) and also ${\gamma _{{\text{th}}}}$ with $\gamma _{{\text{th}}}^{{\text{HD}}} = {(1 + {\gamma _{{\text{th}}}})^2} - 1$, which is obtained by equating the spectral efficiencies of the two schemes, i.e., $\log (1 + {\gamma _{{\text{th}}}}) = \tfrac{1}{2}\log (1 + \gamma _{{\text{th}}}^{{\text{HD}}})$. The latter replacement is made for fair comparison between FD and HD scenarios in terms of outage probability.

\section{Optimization Problem Formulation and Computational Complexity} \label{section:Problem_formulation}

In this section, we study the minimization of the upper bound of system outage probability ${\overline P_{{\text{out}}}}(\mathcal{I},{\mathbf{L}})$, given by \eqref{equation:upper_bound_system_outage_probability}, under various constraints. In particular, the IRS deployment problem consists of two components, namely, the selection of locations for installing IRSs and the determination of IRS sizes (that is, the number of reflecting elements). 

Herein, we consider a \emph{predetermined} and \emph{finite} set of available IRS locations, thus taking into account physical constraints for the IRS positions. This is of great practical interest, since IRSs are usually installed on the facades, walls or ceilings of existing buildings. However, if we are interested in installing IRSs within a bounded continuous area, then this region can be divided into a sufficiently large finite number of distinct points (this technique is known as discretization). As a result, the proposed approach is still applicable. 

As reported in Section \ref{section:System_model}, the IRSs in the set $\mathcal{I}$ are installed only once, during the initial design of the system. After the installation phase, the central controller performs, in each time-slot, the IRS activation between the installed IRSs. 

In this context, the \emph{joint IRS location and size optimization} problem is formulated as follows  
\begin{subequations} \label{equation:initial_problem}
\begin{alignat}{3}
  & \mathop {\min }\limits_{{\mathcal{I},{\mathbf{L}}}} & \quad & {{\overline P}_{{\text{out}}}}(\mathcal{I},{\mathbf{L}}) \triangleq \prod\limits_{i \in \mathcal{I}} {{{\left[ {F_i (\sqrt {{\gamma _{{\text{th}}}}/{\rho _i}} )} \right]}^{{L_i}}}}   \\
  & ~\text{s.t.} & & \mathcal{I} \subseteq \mathcal{N} \\
  & & & {L_n} \in \{ L_n^{\min }, \ldots ,L_n^{\max }\} ,\;\;\forall n \in \mathcal{N} \\
  & & & \left| \mathcal{I} \right| \leq M  \\ \label{equation:initial_constraint_L_tot_max}
  & & &  {L_{{\text{tot}}}}(\mathcal{I},{\mathbf{L}}) \triangleq \sum\limits_{i \in \mathcal{I}} {{L_i}}  \leq L_{{\text{tot}}}^{\max }   \\  \label{equation:initial_constraint_C_tot_max}
  & & &  {C_{{\text{tot}}}}(\mathcal{I},{\mathbf{L}}) \triangleq \sum\limits_{i \in \mathcal{I}} {({c_i} + {\lambda _i}{L_i})}  \leq C_{{\text{tot}}}^{\max }    ,
\end{alignat}
\end{subequations}
where $L_n^{\min },L_n^{\max } \geq 0$ are the minimum and maximum number of reflecting elements of the ${n^{{\text{th}}}}$ IRS, respectively (with $L_n^{\min } \leq L_n^{\max }$). For example, IRS manufacturers may have some restrictions on the production process, while there are space limitations on the area (dimensions) that an IRS can occupy in a specific location/building.\footnote{In addition, $L_n^{\max}$ can be appropriately chosen to ensure that the spatial correlation among the IRS elements is negligible, i.e., $d \geq \lambda/2$, where $d$ is their separation distance and $\lambda$ is the wavelength. For example, let $S_n$ be the maximum rectangular area of the $n^\text{th}$ IRS. Assuming uniform rectangular deployment of IRS elements (with $L_n = {L_n^x}{L_n^y}$ and $d^x = d^y = \lambda/2$), we should guarantee that $(L_n^x-1)(L_n^y-1){(\lambda/2)^2} \leq S_n$. This inequality is satisfied if $L_n^{\max} = 4{S_n}/{\lambda^2}$, because $(L_n^x-1)(L_n^y-1) \leq L_n \leq L_n^{\max}$.} Also, $M \in \{ 0,1, \ldots ,N\} $ is the maximum number of installed IRSs, resulting in an \emph{IRS-cardinality constraint}. Finally, $L_{{\text{tot}}}^{\max },C_{{\text{tot}}}^{\max } \geq 0$ denote the maximum total number of reflecting elements and the maximum total IRS installation cost, respectively. Note that constraint \eqref{equation:initial_constraint_L_tot_max} implicitly imposes an upper bound on the \emph{overall signaling overhead} (channel estimation and feedback), which is required for IRS activation and phase adjustments (see Section \ref{subsection:Implementation}). In addition, for a given $\mathcal{I} \subseteq \mathcal{N}$, the values of ${\{ {L_n}\} _{n \in \mathcal{N}\backslash \mathcal{I}}}$ are ultimately meaningless, since no IRS is installed at these locations. For convenience, Table \ref{table:Math_Symbols} summarizes the mathematical symbols used in optimization throughout the paper.

\begin{table*}[!t]
\caption{List of Mathematical Symbols Used in Optimization} 
\centering
\renewcommand{\arraystretch}{1.8}
\begin{tabular}{|c|c||c|c|}
\hline
\textbf{Symbol} & \textbf{Description} & \textbf{Symbol} & \textbf{Description}\\ \hline
$\mathcal{N} = \{ 1, \ldots ,N\}$ & Set of available locations for installing an IRS & $L_{{\text{tot}}}^{\max}$ & Maximum total number of reflecting elements  \\ \hline
$\mathcal{I} = \{ {i_1}, \ldots ,{i_I}\}$ & Set of finally installed IRSs & \makecell{${C_{{\text{tot}}}}(\mathcal{I},{\mathbf{L}})$ or \\ ${C_{{\text{tot}}}}({\mathbf{x}},{\mathbf{L}})$} & Total IRS installation cost \\ \hline
${x_n}$ & \makecell{Binary (0/1) variable indicating whether an IRS is \\ finally installed at the $n^\text{th}$ location (${x_n} = 1$ iff $n \in \mathcal{I}$) } & $C_{{\text{tot}}}^{\max}$ & Maximum total IRS installation cost \\ \hline
${\mathbf{x}} = {[{x_1}, \ldots ,{x_N}]^ \top }$ & Vector of binary (0/1) variables $\{x_n\}_{n \in \mathcal{N}}$ & $c_n$ & Fixed deployment cost of the $n^{\text{th}}$ IRS \\ \hline
$L_n$ & Number of reflecting elements of the $n^\text{th}$ IRS & $\lambda_n$ & Cost rate (cost-units/element) of the $n^\text{th}$ IRS \\ \hline
${\mathbf{L}} = {[{L_1}, \ldots ,{L_N}]^ \top }$ & Vector of integer variables $\{L_n\}_{n \in \mathcal{N}}$ & $G({\mathbf{x}},{\mathbf{L}})$ & Objective function of discrete problem \eqref{equation:discrete_problem}  \\ \hline
\makecell{${\overline P_{{\text{out}}}}(\mathcal{I},{\mathbf{L}})$ or \\ ${\overline P_{{\text{out}}}}({\mathbf{x}},{\mathbf{L}})$} & Upper bound of system outage probability  & ${G^ \star }$ &  Global minimum of discrete problem \eqref{equation:discrete_problem}  \\ \hline
$L_n^{\min}$, $L_n^{\max}$ & \makecell{Minimum and maximum number of reflecting \\ elements of the $n^{\text{th}}$ IRS, respectively} & $G^\dag$ &  Global minimum of LPR problem \eqref{equation:LPR_problem}, ${G^\dag } \leq {G^ \star}$ \\ \hline
$M$ & \makecell{Maximum number of finally \\ installed IRSs, $0 \leq M \leq N$} & $G'$ &  Objective value obtained from LPR-GA, ${G^ \star} \leq G'$  \\ \hline
\makecell{${L_{{\text{tot}}}}(\mathcal{I},{\mathbf{L}})$ or \\ ${L_{{\text{tot}}}}({\mathbf{x}},{\mathbf{L}})$} & Total number of reflecting elements & $\widetilde G$ &  Objective value achieved by LPR-RA   \\ \hline 
\end{tabular}
\label{table:Math_Symbols}
\end{table*}

\subsection{Transformation into Discrete Optimization Problem}

Now, let us introduce a vector of binary (0/1) variables \linebreak ${\mathbf{x}} = {[{x_1}, \ldots ,{x_N}]^ \top }$ such that, for all $n \in \mathcal{N}$, ${x_n} = 1$ if and only if (iff) $n \in \mathcal{I}$. Subsequently, the set $\mathcal{I}$ is replaced by the vector ${\mathbf{x}}$ in all functions that contained it with a slight abuse of notation. In particular, $\mathcal{I}$ and ${\mathbf{x}}$ are interchangeable because the one can be derived from the other by exploiting their iff-relation. With these in mind, we can make the following observations:\footnote{In order to avoid the undefined quantity ${0^0}$, we assume that ${F_n(\sqrt {{\gamma _{{\text{th}}}}/{\rho _n}} )} > 0$, for all $n \in \mathcal{N}$.} 1) ${\overline P_{{\text{out}}}}(\mathcal{I},{\mathbf{L}}) = \prod_{n \in \mathcal{N}} {{{\left[ {F_n(\sqrt {{\gamma _{{\text{th}}}}/{\rho _n}} )} \right]}^{{x_n}{L_n}}}}$, 2) $\left| \mathcal{I} \right| = \sum_{n \in \mathcal{N}} {{x_n}}$, 3) ${L_{{\text{tot}}}}(\mathcal{I},{\mathbf{L}}) = \sum_{n \in \mathcal{N}} {{x_n}{L_n}}$ and 4) ${C_{{\text{tot}}}}(\mathcal{I},{\mathbf{L}}) = \sum_{n \in \mathcal{N}} {({c_n} + {\lambda _n}{L_n}){x_n}}$. Therefore, problem \eqref{equation:initial_problem} can be written as follows 
\begin{subequations} 
\begin{alignat}{3}
  & \mathop {\min }\limits_{{{\mathbf{x}},{\mathbf{L}}}} & \quad & {{\overline P}_{{\text{out}}}}({\mathbf{x}},{\mathbf{L}}) \triangleq \prod\limits_{n \in \mathcal{N}} {{{\left[ {F_n(\sqrt {{\gamma _{{\text{th}}}}/{\rho _n}} )} \right]}^{{x_n}{L_n}}}}    \\
  & ~\text{s.t.} & & {x_n} \in \{ 0,1\} ,\;\;\forall n \in \mathcal{N} \\
  & & & {L_n} \in \{ L_n^{\min }, \ldots ,L_n^{\max }\} ,\;\;\forall n \in \mathcal{N} \\
  & & &  \sum\limits_{n \in \mathcal{N}} {{x_n}}  \leq M \\ 
  & & &  {L_{{\text{tot}}}}({\mathbf{x}},{\mathbf{L}}) \triangleq \sum\limits_{n \in \mathcal{N}} {{x_n}{L_n}}  \leq L_{{\text{tot}}}^{\max }   \\ 
  & & &   {C_{{\text{tot}}}}({\mathbf{x}},{\mathbf{L}}) \triangleq \sum\limits_{n \in \mathcal{N}} {({c_n} + {\lambda _n}{L_n}){x_n}} \leq C_{{\text{tot}}}^{\max }   ,
\end{alignat}
\end{subequations}
where ${\mathbf{x}}$, ${\mathbf{L}}$ are the decision/optimization variables. 

Since $\log ( \cdot )$ is a monotonically increasing function, we can replace ${{\overline P}_{{\text{out}}}}({\mathbf{x}},{\mathbf{L}})$ with its logarithm, without altering the set of optimal solutions. Hence, we obtain the equivalent \emph{discrete optimization problem} 
\begin{subequations} \label{equation:discrete_problem}
\begin{alignat}{3}
  & \mathop {\min }\limits_{{{\mathbf{x}},{\mathbf{L}}}} & \quad & G({\mathbf{x}},{\mathbf{L}}) \triangleq \log \left( {{{\overline P}_{{\text{out}}}}({\mathbf{x}},{\mathbf{L}})} \right) = \sum\limits_{n \in \mathcal{N}} {{\beta _n}({x_n}{L_n})}     \\
  & ~\text{s.t.} & & {x_n} \in \{ 0,1\} ,\;\;\forall n \in \mathcal{N} \\
  & & & {L_n} \in \{ L_n^{\min }, \ldots ,L_n^{\max }\} ,\;\;\forall n \in \mathcal{N} \\ \label{equation:constraint_M}
  & & &  \sum\limits_{n \in \mathcal{N}} {{x_n}}  \leq M \\ \label{equation:constraint_L_tot_max}
  & & &  \sum\limits_{n \in \mathcal{N}} {{x_n}{L_n}}  \leq L_{{\text{tot}}}^{\max }   \\ \label{equation:constraint_C_tot_max}
  & & &   \sum\limits_{n \in \mathcal{N}} {{c_n}{x_n}}  + \sum\limits_{n \in \mathcal{N}} {{\lambda _n}({x_n}{L_n})} \leq C_{{\text{tot}}}^{\max }   ,
\end{alignat}
\end{subequations}
where ${\beta _n} = \log \left( {F_n(\sqrt {{\gamma _{{\text{th}}}}/{\rho _n}} )} \right) \leq 0$ for all $n \in \mathcal{N}$. Throughout the paper, $({{\mathbf{x}}^ \star },{{\mathbf{L}}^ \star })$ and ${G^ \star } = G({{\mathbf{x}}^ \star },{{\mathbf{L}}^ \star })$ denote an optimal solution and the global minimum of problem \eqref{equation:discrete_problem}, respectively. As we will see later, this problem is rather unlikely to be globally solved in polynomial time due to its discrete (and, thus, nonconvex) structure. 

\vspace{2mm}
\begin{remark}[Feasibility]  \label{remark:guaranteed_feasibility}
Optimization problem \eqref{equation:discrete_problem} is \emph{always} feasible, since the solution $({\mathbf{x}},{\mathbf{L}}) = ({{\mathbf{0}}_N},{{\mathbf{L}}^{\min }})$, where ${{\mathbf{L}}^{\min }} = {[L_1^{\min }, \ldots ,L_N^{\min }]^ \top }$, satisfies all constraints. 
\end{remark}

\subsection{NP-Hardness}

Afterwards, we examine the computational complexity of finding a (globally) optimal solution to problem \eqref{equation:discrete_problem}. 

\vspace{2mm}
\begin{theorem}[NP-hardness] \label{theorem:NP_hardness} 
Assume that the functions $\{ F_n( \cdot ) \}_{n \in \mathcal{N}}$, defined in \eqref{equation:CDF_single_element}, are continuous and (strictly) increasing. Then, the discrete optimization problem \eqref{equation:discrete_problem} is NP-hard. 
\end{theorem}
\vspace{2mm}

\renewcommand{\IEEEQED}{\IEEEQEDopen}
\begin{IEEEproof}
See Appendix \ref{appendix:NP_hardness}.    
\end{IEEEproof}

\section{Optimization Algorithms} \label{section:Optimization_algorithms}

Subsequently, we present the exhaustive-search (or brute-force) technique, a linear-programming relaxation (LPR), a greedy method as well as a randomized algorithm with complexity analysis for each one. The solution of LPR plays a central role in the design of greedy and randomized algorithms.

\subsection{Exhaustive-Enumeration Algorithm} \label{subsection:Exhaustive}

The exhaustive-enumeration method checks all possible solutions and selects that with the minimum objective value satisfying all constraints. In particular, for all subsets $\mathcal{I}$ of $\mathcal{N}$ with cardinality at most $M$, the algorithm examines all arrangements of reflecting elements. More specifically, for a given subset $\mathcal{I}$, the number of arrangements of reflecting elements is $\prod_{i \in \mathcal{I}} {(L_i^{\max } - L_i^{\min } + 1)}$, because the decision variable ${L_i}$ can take $(L_i^{\max } - L_i^{\min } + 1)$ distinct values. In addition, for each arrangement, the algorithm requires $\Theta \left( {\left| \mathcal{I} \right|} \right)$ time to compute the objective value, $G({\mathcal{I}},{\mathbf{L}}) = \sum_{i \in \mathcal{I}} {{\beta _i}{L_i}}$, and check the feasibility of constraints ${L_{{\text{tot}}}}(\mathcal{I},{\mathbf{L}}) \leq L_{{\text{tot}}}^{\max }$, ${C_{{\text{tot}}}}(\mathcal{I},{\mathbf{L}}) \leq C_{{\text{tot}}}^{\max }$. Therefore, its overall runtime is $\Theta \left( \sum_{\begin{subarray}{l} \mathcal{I} \subseteq \mathcal{N} \\ \left| \mathcal{I} \right| \leq M \end{subarray}}  {\left| \mathcal{I} \right| \prod_{i \in \mathcal{I}} {R_i} } \right)$, where ${R_n} = L_n^{\max } - L_n^{\min } + 1$ for all $n \in \mathcal{N}$. 

It is not difficult to conclude that the algorithm has \linebreak \emph{exponential complexity} in terms of the size of the problem. Let us suppose that $R_n = R \geq 1$ for all $n \in \mathcal{N}$. In this case, the algorithm requires $\Theta \left( \sum_{\begin{subarray}{l} \mathcal{I} \subseteq \mathcal{N} \\ \left| \mathcal{I} \right| \leq M \end{subarray}}  {\left| \mathcal{I} \right| {R^{\left| \mathcal{I} \right|}} } \right) = \Omega \left( \sum_{\begin{subarray}{l} \mathcal{I} \subseteq \mathcal{N} \\ \left| \mathcal{I} \right| = M \end{subarray}}  {\left| \mathcal{I} \right| {R^{\left| \mathcal{I} \right|}} } \right) = \Omega \left( {\binom{N}{M}} M {R^M} \right)$ arithmetic operations to find the global minimum. Furthermore, if $M = \left\lceil {N/2} \right\rceil$, then its complexity becomes $\Omega \left( {{2^N}{\sqrt N} {R^{N/2}}} \right)$, since $\binom{N}{\left\lceil {N/2} \right\rceil} = \Theta \left( \frac{2^N}{\sqrt N} \right)$ and $\left\lceil {N/2} \right\rceil \geq N/2$. 

Ultimately, albeit achieving a globally optimal solution, the exhaustive-enumeration algorithm has extremely high complexity and is therefore impractical.

\subsection{Lower Bound Using Linear-Programming Relaxation} \label{subsection:LPR}

Despite the difficulty of computing the global minimum, we will show how to efficiently compute (in polynomial-time) a lower bound of the optimum value ${G^ \star }$. Firstly, by using auxiliary decision variables ${\mathbf{z}} = {[{z_1}, \ldots ,{z_N}]^ \top }$, problem \eqref{equation:discrete_problem} can be equivalently written in the following form 
\begin{subequations} \label{equation:pre2_LPR_problem}
\begin{alignat}{3}
  & \mathop {\min }\limits_{{{\mathbf{x}},{\mathbf{L}},{\mathbf{z}}}} & \quad & \sum\limits_{n \in \mathcal{N}} {{\beta _n}{z_n}}  \\
  & ~\text{s.t.} & & {x_n} \in \{ 0,1\} ,\;\;\forall n \in \mathcal{N} \\
  & & & {L_n} \in \{ L_n^{\min }, \ldots ,L_n^{\max }\} ,\;\;\forall n \in \mathcal{N} \\
  & & & {z_n} = {x_n}{L_n},\;\;\forall n \in \mathcal{N} \\
  & & &  \sum\limits_{n \in \mathcal{N}} {{x_n}}  \leq M \\ 
  & & &  \sum\limits_{n \in \mathcal{N}} {{z_n}}  \leq L_{{\text{tot}}}^{\max }   \\ 
  & & &  \sum\limits_{n \in \mathcal{N}} {{c_n}{x_n}}  + \sum\limits_{n \in \mathcal{N}} {{\lambda _n}{z_n}} \leq C_{{\text{tot}}}^{\max }   .
\end{alignat} 
\end{subequations}
Secondly, by relaxing the integer/discrete constraints, \linebreak ${x_n} \in \{ 0,1\}$ and ${L_n} \in \{ L_n^{\min }, \ldots ,L_n^{\max }\} $, we have  
\begin{subequations} \label{equation:pre_LPR_problem}
\begin{alignat}{3}
  & \mathop {\min }\limits_{{{\mathbf{x}},{\mathbf{L}},{\mathbf{z}}}} & \quad & \sum\limits_{n \in \mathcal{N}} {{\beta _n}{z_n}}  \\
  & ~\text{s.t.} & & 0 \leq {x_n} \leq 1,\;\;\forall n \in \mathcal{N} \\
  & & & L_n^{\min } \leq {L_n} \leq L_n^{\max },\;\;\forall n \in \mathcal{N} \\
  & & & {z_n} = {x_n}{L_n},\;\;\forall n \in \mathcal{N} \\
  & & &  \sum\limits_{n \in \mathcal{N}} {{x_n}}  \leq M \\ 
  & & &  \sum\limits_{n \in \mathcal{N}} {{z_n}}  \leq L_{{\text{tot}}}^{\max }   \\ 
  & & &  \sum\limits_{n \in \mathcal{N}} {{c_n}{x_n}}  + \sum\limits_{n \in \mathcal{N}} {{\lambda _n}{z_n}} \leq C_{{\text{tot}}}^{\max }   .
\end{alignat}
\end{subequations}
Notice that this problem is nonlinear due to the equality constraints ${z_n} = {x_n}{L_n}$. In order to obtain a linear problem, we apply further relaxation by replacing the set of constraints $L_n^{\min } \leq {L_n} \leq L_n^{\max }$ and ${z_n} = {x_n}{L_n}$ with the linear constraints $L_n^{\min }{x_n} \leq {z_n} \leq L_n^{\max }{x_n}$. In this way, we can remove the decision variable ${\mathbf{L}}$ and formulate the following \emph{linear-programming relaxation (LPR)} problem  
\begin{subequations} \label{equation:LPR_problem}
\begin{alignat}{3}
  & \mathop {\min }\limits_{{{\mathbf{x}},{\mathbf{z}}}} & \quad & \sum\limits_{n \in \mathcal{N}} {{\beta _n}{z_n}}  \\
  & ~\text{s.t.} & & 0 \leq {x_n} \leq 1,\;\;\forall n \in \mathcal{N} \\
  & & & L_n^{\min }{x_n} \leq {z_n} \leq L_n^{\max }{x_n},\;\;\forall n \in \mathcal{N} \\
  & & &  \sum\limits_{n \in \mathcal{N}} {{x_n}}  \leq M \\ 
  & & &  \sum\limits_{n \in \mathcal{N}} {{z_n}}  \leq L_{{\text{tot}}}^{\max }   \\ 
  & & &  \sum\limits_{n \in \mathcal{N}} {{c_n}{x_n}}  + \sum\limits_{n \in \mathcal{N}} {{\lambda _n}{z_n}} \leq C_{{\text{tot}}}^{\max }   .
\end{alignat}
\end{subequations}

Note that the guaranteed feasibility of problem \eqref{equation:discrete_problem} (see Remark \ref{remark:guaranteed_feasibility}) implies the feasibility of problems \eqref{equation:pre2_LPR_problem}, \eqref{equation:pre_LPR_problem} and \eqref{equation:LPR_problem}. In what follows, $({{\mathbf{x}}^\dag },{{\mathbf{z}}^\dag })$ and ${G^\dag } = \sum_{n \in \mathcal{N}} {{\beta _n}z_n^\dag } $ denote an optimal solution and the global minimum of the LPR problem \eqref{equation:LPR_problem}, respectively. Obviously, ${G^\dag } \leq {G^ \star }$, that is, ${G^\dag }$ is a lower bound of ${G^ \star }$.

Finally, given that the linear problem \eqref{equation:LPR_problem} has $V = 2N = \Theta (N)$ decision variables and $U = 4N + 3 = \Theta (N)$ constraints, a globally optimal solution can be computed in $O({(U + V)^{1.5}}{V^2}) = O({N^{3.5}})$ time using an interior-point method \cite{Ben-Tal_Nemirovski}.

\subsection{Deterministic Greedy Algorithm}

Now, we are ready to develop a heuristic algorithm of polynomial complexity to obtain a feasible solution for the discrete problem \eqref{equation:discrete_problem}. This procedure is given in Algorithm \ref{algorithm:LPR-GA} and is called \emph{LPR-based greedy algorithm (LPR-GA)}.

\begin{algorithm}[!t]
\caption{LPR-based Greedy Algorithm (LPR-GA)} \label{algorithm:LPR-GA}
\small
\begin{algorithmic}[1]
\Require $N$, ${\boldsymbol{\beta }} = {[{\beta _1}, \ldots ,{\beta _N}]^ \top }$, ${{\mathbf{L}}^{\min }} = {[L_1^{\min }, \ldots ,L_N^{\min }]^ \top }$, \begin{flushleft}~~~~$\,{{\mathbf{L}}^{\max }} = {[L_1^{\max }, \ldots ,L_N^{\max }]^ \top }$, $M$, $L_{{\text{tot}}}^{\max }$, ${\mathbf{c}} = {[{c_1}, \ldots ,{c_N}]^ \top }$, \end{flushleft} \break \begin{flushleft}~~~~$\,{\boldsymbol{\lambda }} = {[{\lambda _1}, \ldots ,{\lambda _N}]^ \top }$, $C_{{\text{tot}}}^{\max }$ \end{flushleft} 
\Ensure A feasible solution $({\mathbf{x'}},{\mathbf{L'}})$ of discrete problem \eqref{equation:discrete_problem}
\State Solve the LPR problem \eqref{equation:LPR_problem} to obtain an optimal \break solution $({{\mathbf{x}}^\dag },{{\mathbf{z}}^\dag })$.
\State $\triangleright$ \textit{Computation of} ${\mathbf{L'}} = {[{L'_1}, \ldots ,{L'_N}]^ \top }$ 
\ForAll{$n \in \mathcal{N}$}
	\If{$x_n^\dag \ne 0$}
		\State ${L'_n} : = \operatorname{round} \left( {z_n^\dag /x_n^\dag } \right)$  
	\Else
    	\State ${L'_n} : = \operatorname{round} \left( {\tfrac{1}{2}(L_n^{\min } + L_n^{\max })} \right)$
	\EndIf	
\EndFor
\State $\triangleright$ \textit{Computation of} ${\mathbf{x'}} = {[{x'_1}, \ldots ,{x'_N}]^ \top }$
\State Sort the entries of ${{\mathbf{x}}^\dag }$ in descending order. Let $({\sigma _1}, \ldots ,{\sigma _N}) \in {\Sigma_\mathcal{N}}$ be their order after sorting, where ${\Sigma _\mathcal{N}}$ is the set of all permutations of $\mathcal{N}$, therefore \break $x_{{\sigma _1}}^\dag  \geq  \cdots  \geq x_{{\sigma _N}}^\dag$.  
\State $m: = 1$, ${L_{{\text{tot}}}}: = 0$, ${C_{{\text{tot}}}}: = 0$, ${\mathbf{x'}}: = {{\mathbf{0}}_N}$ 
\While{$(m \leq M) \wedge ({L_{{\text{tot}}}} \leq L_{{\text{tot}}}^{\max }) \wedge ({C_{{\text{tot}}}} \leq C_{{\text{tot}}}^{\max })$}
	\State $i: = {\sigma _m}$, ${x'_i}: = 1$
	\State ${L_{{\text{tot}}}}: = {L_{{\text{tot}}}} + {L'_i}$, ${C_{{\text{tot}}}}: = {C_{{\text{tot}}}} + {c_i} + {\lambda_i}{L'_i}$
	\State $m: = m + 1$
\EndWhile
\If{$({L_{{\text{tot}}}} > L_{{\text{tot}}}^{\max }) \vee ({C_{{\text{tot}}}} > C_{{\text{tot}}}^{\max })$}
	\State ${x'_i}: = 0$
\EndIf
\State \textbf{return} $({\mathbf{x'}},{\mathbf{L'}})$
\end{algorithmic}
\end{algorithm}

First, the proposed algorithm applies \emph{deterministic rounding}, using the solution of LPR (step 1), in order to compute the decision variable ${\mathbf{L'}}$ (steps 3--9): 
\begin{equation}
{L'_n} = \left\{ \begin{gathered}
  \operatorname{round} \left( {z_n^\dag /x_n^\dag } \right),\; {\text{if}}\;x_n^\dag  \ne 0 \hfill \\
  \operatorname{round} \left( {\tfrac{1}{2}(L_n^{\min } + L_n^{\max })} \right),\; {\text{otherwise}} \hfill \\ 
\end{gathered}  \right.,\; \forall n \in \mathcal{N} .
\end{equation}
Observe that, if $x_n^\dag \ne 0$, then $L_n^{\min } \leq z_n^\dag /x_n^\dag  \leq L_n^{\max }$ (due to the feasibility of LPR problem) and therefore $\operatorname{round} \left( {z_n^\dag /x_n^\dag } \right) \in \{ L_n^{\min }, \ldots ,L_n^{\max }\} $. Also, the same holds for $\operatorname{round} \left( {\tfrac{1}{2}(L_n^{\min } + L_n^{\max })} \right)$ in case of $x_n^\dag  = 0$. In other words, the above deterministic rounding guarantees that  ${L'_n} \in \{ L_n^{\min }, \ldots ,L_n^{\max }\} $ for all $n \in \mathcal{N}$. 

Concerning the computation of $\mathbf{x'}$, the proposed algorithm sorts the entries of ${{\mathbf{x}}^\dag } = {[x_1^\dag , \ldots ,x_N^\dag ]^ \top } \in {[0,1]^N}$ in descending order (step 11). Then, by starting from the zero vector, it successively selects IRS locations (based on their sorting) until the violation of at least one of the constraints: $\sum\nolimits_{n \in \mathcal{N}} {{x_n}} \leq M$, ${L_{{\text{tot}}}}({\mathbf{x}},{\mathbf{L}}) \leq L_{{\text{tot}}}^{\max }$ and ${C_{{\text{tot}}}}({\mathbf{x}},{\mathbf{L}}) \leq C_{{\text{tot}}}^{\max }$ (steps 12--17). Finally, it removes the last selected IRS location if ${L_{{\text{tot}}}} > L_{{\text{tot}}}^{\max }$ or ${C_{{\text{tot}}}} > C_{{\text{tot}}}^{\max }$ (steps 18--20); note that the IRS-cardinality constraint is automatically satisfied due to the construction of the while-loop, so there is no need to check it in step 18. 

Obviously, the solution $({\mathbf{x'}},{\mathbf{L'}})$ returned by Algorithm \ref{algorithm:LPR-GA} is guaranteed to be feasible for problem \eqref{equation:discrete_problem}, thus ${G^ \star } \leq G' \triangleq G({\mathbf{x'}},{\mathbf{L'}})$, i.e., $G'$ is an upper bound of ${G^ \star }$.  

\vspace{2mm}
\begin{remark}[A posteriori performance guarantee]
It is possible to provide an approximation guarantee after the termination of Algorithm \ref{algorithm:LPR-GA} using the already obtained solution of LPR, that is, $0 \leq G' - {G^ \star } \leq G' - {G^\dag }$.
\end{remark}
\vspace{2mm}

Regarding the complexity of Algorithm \ref{algorithm:LPR-GA}, the LPR problem can be solved in $O({N^{3.5}})$ time, the computation of ${\mathbf{L'}}$ requires $\Theta (N)$ time, while the computation of ${\mathbf{x'}}$ requires $O(N\log N + N) = O(N\log N)$ arithmetic operations in total. Hence, the overall complexity of LPR-GA is $O({N^{3.5}} + N + N\log N) = O({N^{3.5}})$. In other words, it has the same asymptotic complexity (up to a constant) as the LPR problem.

\subsection{Randomized Approximation Algorithm}

Since problem \eqref{equation:discrete_problem} is NP-hard, a polynomial-time algorithm for computing its global optimum cannot exist, unless P=NP. However, we can use \emph{randomized rounding}, a powerful technique, in order to achieve an \emph{approximate solution with high probability}. 

Afterwards, we present an efficient (i.e., polynomial-time) approximation algorithm that finds \emph{provably near-optimal solutions}. This method is shown in Algorithm \ref{algorithm:LPR-RA} and is referred to as \emph{LPR-based randomized algorithm (LPR-RA)}.\footnote{Note that the function $\operatorname{rand} (0,1)$ in steps 4 and 12 returns a random number uniformly distributed in the interval $[0,1]$. Also, the function $\operatorname{randi} (L_n^{\min },L_n^{\max })$ in step 19 returns a random integer uniformly distributed in the set $\{ L_n^{\min }, \ldots ,L_n^{\max }\}$. The outputs of different function calls are \emph{independent}.}

\begin{algorithm}[!t]
\caption{LPR-based Randomized Algorithm (LPR-RA)} \label{algorithm:LPR-RA}
\small
\begin{algorithmic}[1]
\Require $N$, ${\boldsymbol{\beta }} = {[{\beta _1}, \ldots ,{\beta _N}]^ \top }$, ${{\mathbf{L}}^{\min }} = {[L_1^{\min }, \ldots ,L_N^{\min }]^ \top }$, \begin{flushleft}~~~~$\,{{\mathbf{L}}^{\max }} = {[L_1^{\max }, \ldots ,L_N^{\max }]^ \top }$, $M$, $L_{{\text{tot}}}^{\max }$, ${\mathbf{c}} = {[{c_1}, \ldots ,{c_N}]^ \top }$, \end{flushleft} \break \begin{flushleft}~~~~$\,{\boldsymbol{\lambda }} = {[{\lambda _1}, \ldots ,{\lambda _N}]^ \top }$, $C_{{\text{tot}}}^{\max }$ \end{flushleft} 
\Ensure An approximate solution $({\mathbf{\widetilde x}},\widetilde {\mathbf{L}})$ of discrete problem \eqref{equation:discrete_problem}
\State Solve the LPR problem \eqref{equation:LPR_problem} to obtain an optimal \break solution $({{\mathbf{x}}^\dag },{{\mathbf{z}}^\dag })$.
\ForAll{$n \in \mathcal{N}$}
	\State $\triangleright$ \textit{Computation of} ${\widetilde x_n}$
	\State $r: = \operatorname{rand}(0,1)$
	\If{$r \leq x_n^\dag$}        \hfill $\triangleright$ \textit{with probability} $x_n^\dag$
		\State ${\widetilde x_n}: = 1$
	\Else       \hfill $\triangleright$ \textit{with probability} $1 - x_n^\dag$
		\State ${\widetilde x_n}: = 0$
	\EndIf
	\State $\triangleright$ \textit{Computation of} ${\widetilde L_n}$
	\If{$x_n^\dag \ne 0$}
		\State $s: = \operatorname{rand}(0,1)$, $L_n^\dag : = z_n^\dag /x_n^\dag$
		\If{$s \leq \operatorname{frac} (L_n^\dag )$}      \hfill $\triangleright$ \textit{with probability} $\operatorname{frac} (L_n^\dag )$
			\State ${\widetilde L_n}: = \left\lfloor {L_n^\dag } \right\rfloor  + 1$
		\Else      \hfill $\triangleright$ \textit{with probability} $1 - \operatorname{frac} (L_n^\dag )$
			\State ${\widetilde L_n}: = \left\lfloor {L_n^\dag } \right\rfloor$
		\EndIf
	\Else
		\State ${\widetilde L_n}: = \operatorname{randi} (L_n^{\min },L_n^{\max })$
	\EndIf
\EndFor
\State \textbf{return} $({\mathbf{\widetilde x}},\widetilde {\mathbf{L}})$ 
\end{algorithmic}
\end{algorithm}

First of all, Algorithm \ref{algorithm:LPR-RA} finds an optimal solution $({{\mathbf{x}}^\dag },{{\mathbf{z}}^\dag })$ to the LPR problem in step 1, and then employs randomization (steps 2--21) to compute an approximate solution $({\mathbf{\widetilde x}},\widetilde {\mathbf{L}})$ according to the following probabilistic rules, for all $n \in \mathcal{N}$, 
\begin{equation} \label{equation:x_approx}
{\widetilde x_n} = \left\{ \begin{gathered}
  1,\;\; {\text{with probability}}\ x_n^\dag  \hfill \\
  0,\;\; {\text{with probability}}\ 1 - x_n^\dag  \hfill \\ 
\end{gathered}  \right. 
\end{equation}
and
\begin{equation} \label{equation:L_approx_a}
{\widetilde L_n} = \left\{ \begin{gathered}
  \left\lfloor {L_n^\dag } \right\rfloor  + 1,\;\; {\text{with probability}}\ \operatorname{frac} (L_n^\dag ) \hfill \\
  \left\lfloor {L_n^\dag } \right\rfloor ,\;\; {\text{with probability}}\ 1 - \operatorname{frac} (L_n^\dag ) \hfill \\ 
\end{gathered}  \right.,\;\; {\text{if}}\ x_n^\dag \ne 0 
\end{equation}
or
\begin{equation} \label{equation:L_approx_b}
{\widetilde L_n} \sim \operatorname{Uniform} \left( {\{ L_n^{\min }, \ldots ,L_n^{\max }\} } \right),\;\; {\text{if}}\ x_n^\dag = 0  ,
\end{equation}
where $L_n^\dag  = z_n^\dag /x_n^\dag$ defined whenever $x_n^\dag \ne 0$. For a given LPR solution $({{\mathbf{x}}^\dag },{{\mathbf{z}}^\dag })$, the random variables $\{{\widetilde x_n},{\widetilde L_n}\}_{n \in \mathcal{N}}$ are \emph{independent}. Also, we define the random variable ${\widetilde z_n} = {\widetilde x_n}{\widetilde L_n}$ for all $n \in \mathcal{N}$.

In case $x_n^\dag  \ne 0$, observe that $L_n^{\min } \leq L_n^\dag  \leq L_n^{\max }$ (because $L_n^{\min }x_n^\dag  \leq z_n^\dag  \leq L_n^{\max }x_n^\dag $), therefore rule \eqref{equation:L_approx_a} implies that ${\widetilde L_n} \in \{ L_n^{\min }, \ldots ,L_n^{\max }\} $. Also, the same holds when $x_n^\dag  = 0$ due to \eqref{equation:L_approx_b}. Hence, the probabilistic rules \eqref{equation:x_approx}--\eqref{equation:L_approx_b} ensure that the approximate solution returned by LPR-RA, $({\mathbf{\widetilde x}},\widetilde {\mathbf{L}})$, satisfies the integer/discrete constraints automatically, i.e., ${\widetilde x_n} \in \{ 0,1\} $ and ${\widetilde L_n} \in \{ L_n^{\min }, \ldots ,L_n^{\max }\} $ for every $n \in \mathcal{N}$. 

Now, it remains to answer how close is the achieved objective value $\widetilde G \triangleq G({\mathbf{\widetilde x}},\widetilde {\mathbf{L}})$ to the global minimum ${G^ \star }$, and whether or not the last three constraints, \eqref{equation:constraint_M}, \eqref{equation:constraint_L_tot_max} and \eqref{equation:constraint_C_tot_max}, are satisfied. Of course, there is no absolute/deterministic answer to these type of questions since ${\mathbf{\widetilde x}}$ and $\widetilde {\mathbf{L}}$ are random variables. Nevertheless, we can provide some \emph{(a priori) probabilistic guarantees} on the algorithm's performance.

\vspace{2mm}
\begin{theorem}[Expectation guarantees] \label{theorem:Expectation_guarantees} 
The solution $({\mathbf{\widetilde x}},\widetilde {\mathbf{L}})$ of LPR-RA has the following properties:
\begin{equation} \label{equation:expectation_guarantee_objective}
\mathbb{E}\left( {G({\mathbf{\widetilde x}},\widetilde {\mathbf{L}})} \right) = {G^\dag } ,
\end{equation}
\begin{equation} \label{equation:expectation_guarantee_M}
\mathbb{E}\left( {\sum\limits_{n \in \mathcal{N}} {{{\widetilde x}_n}} } \right) = \sum\limits_{n \in \mathcal{N}} {x_n^\dag } \leq M  ,
\end{equation}
\begin{equation} \label{equation:expectation_guarantee_L_tot_max}
\mathbb{E}\left( {{L_{{\text{tot}}}}({\mathbf{\widetilde x}},\widetilde {\mathbf{L}})} \right) = \sum\limits_{n \in \mathcal{N}} {z_n^\dag }  \leq L_{{\text{tot}}}^{\max }  ,
\end{equation}
\begin{equation} \label{equation:expectation_guarantee_C_tot_max}
\mathbb{E}\left( {{C_{{\text{tot}}}}({\mathbf{\widetilde x}},\widetilde {\mathbf{L}})} \right) = \sum\limits_{n \in \mathcal{N}} {{c_n}x_n^\dag }  + \sum\limits_{n \in \mathcal{N}} {{\lambda _n}z_n^\dag }  \leq C_{{\text{tot}}}^{\max }  .
\end{equation}
\end{theorem}
\vspace{2mm}

\begin{IEEEproof}
See Appendix \ref{appendix:Expectation_guarantees}.
\end{IEEEproof}
\vspace{2mm}

In other words, the solution $({\mathbf{\widetilde x}},\widetilde {\mathbf{L}})$ satisfies \emph{in expectation} the three constraints \eqref{equation:constraint_M}--\eqref{equation:constraint_C_tot_max} and its objective value is \emph{in expectation} equal to that of LPR.  

Afterwards, we give some deviation guarantees using \emph{concentration inequalities}, that is, probability bounds on how close a random variable is from some value (typically, its expectation).

\vspace{2mm}
\begin{theorem}[Deviation guarantees] \label{theorem:Deviation_guarantees} 
Let ${\Delta _0} = \sum_{n \in \mathcal{N}} {{{({\beta _n}L_n^{\max })}^2}} $, ${\Delta _1} = N$ $(\geq \! 1)$, ${\Delta _2} = \sum_{n \in \mathcal{N}} {{{(L_n^{\max })}^2}} $ and ${\Delta _3} = \sum_{n \in \mathcal{N}} {{{({c_n} + {\lambda _n}L_n^{\max })}^2}} $ with ${\Delta _0},{\Delta _2},{\Delta _3} > 0$. Then, the solution $({\mathbf{\widetilde x}},\widetilde {\mathbf{L}})$ achieved by Algorithm \ref{algorithm:LPR-RA} satisfies the following probabilistic conditions:
\begin{equation} \label{equation:deviation_guarantees}
\Pr ({\mathcal{E}_k}) \geq 1 - {\xi},\;\; \forall k \in \{ 0,1,2,3\}  , 
\end{equation}
where ${\xi} = {(N + 1)^{ - 2}} \in (0,1/4]$ and the events $\{ {\mathcal{E}_k}\} _{k = 0}^3$ are defined by 
\begin{equation} \label{equation:event_0}
{\mathcal{E}_0} = \left\{ {G({\mathbf{\widetilde x}},\widetilde {\mathbf{L}}) \leq {G^ \star } + {\epsilon_0}} \right\}   ,
\end{equation}
\begin{equation} \label{equation:event_1}
{\mathcal{E}_1} = \left\{ {\sum\limits_{n \in \mathcal{N}} {{{\widetilde x}_n}}  \leq M + {\epsilon_1}} \right\}  ,
\end{equation}
\begin{equation} \label{equation:event_2}
{\mathcal{E}_2} = \left\{ {{L_{{\text{tot}}}}({\mathbf{\widetilde x}},\widetilde {\mathbf{L}}) \leq L_{{\text{tot}}}^{\max } + {\epsilon_2}} \right\}  ,
\end{equation}
\begin{equation}  \label{equation:event_3}
{\mathcal{E}_3} = \left\{ {{C_{{\text{tot}}}}({\mathbf{\widetilde x}},\widetilde {\mathbf{L}}) \leq C_{{\text{tot}}}^{\max } + {\epsilon_3}} \right\}  ,
\end{equation}
with ${\epsilon_k} = \sqrt {{\Delta_k}\log (N + 1)} > 0$ for all $k \in \{ 0,1,2,3\}$.

Furthermore, the approximate solution $({\mathbf{\widetilde x}},\widetilde {\mathbf{L}})$ satisfies the following inequalities (which are called \emph{overall deviation guarantees}): 
\begin{equation} \label{equation:overall_guarantee_constraints}
\Pr \left( {\bigcap\limits_{k = 1}^3 {{\mathcal{E}_k}} } \right) \geq 1 - \xi'  ,
\end{equation}
\begin{equation} \label{equation:overall_guarantee_objective_constraints}
\Pr \left( {\bigcap\limits_{k = 0}^3 {{\mathcal{E}_k}} } \right) \geq 1 - \xi''  ,
\end{equation}
where $\xi' = 3 \xi  = 3{(N + 1)^{ - 2}} \in (0,3/4]$ and $\xi'' = 4 \xi  = 4{(N + 1)^{ - 2}} \in (0,1]$.
\end{theorem}
\vspace{2mm}

\begin{IEEEproof}
See Appendix \ref{appendix:Deviation_guarantees}.
\end{IEEEproof}
\vspace{2mm}

In essence, $\epsilon_0$ quantifies the deviation from the global optimum, while $\{ {\epsilon_k}\} _{k = 1}^3$ express the violation tolerance for each constraint. Moreover, \eqref{equation:overall_guarantee_constraints} has to do with the satisfiability of the three constraints \eqref{equation:constraint_M}--\eqref{equation:constraint_C_tot_max}, while \eqref{equation:overall_guarantee_objective_constraints} includes additionally the objective value. 

Notice that ${\xi} = {\xi}(N) = o(1)$,  $\xi ' = \xi '(N) = o(1)$ and $\xi '' = \xi ''(N) = o(1)$.\footnote{Here, the asymptotic term $o(1)$ is defined for $N \to \infty $, for example, $\xi  = \xi (N) = o(1)$ means that ${\lim _{N \to \infty }}\xi  = 0$.} Therefore, an approximate solution satisfying any nonempty subset of $\{ {\mathcal{E}_k}\} _{k = 0}^3$ can be found \emph{with high probability}, i.e., with probability $1 - o(1)$. However, observe that there is an inherent tradeoff between the probability bound $1 - \xi$ and a deviation tolerance $\epsilon_k$, that is, both of them increase with $N$ but with different growth rates.

\vspace{2mm}
\begin{remark}[Performance improvement] \label{remark:Performance_improvement}
In order to enhance the algorithm's performance (i.e., achieve better probability bounds), we can use multiple \emph{i.i.d. rounding trials} at the cost of increased computational complexity. For example, let $T$ be the number of rounding trials and $({{\mathbf{\widetilde x}}^{(t)}},{\widetilde {\mathbf{L}}^{(t)}})$ be the approximate solution obtained in rounding trial $t \in \{ 1, \ldots ,T\} $. Also, for every $k \in \{ 1,2,3\} $, let $\mathcal{E}_k^{(t)}$ be the event ${\mathcal{E}_k}$ (see \eqref{equation:event_1}--\eqref{equation:event_3}) with $({\mathbf{\widetilde x}},\widetilde {\mathbf{L}})$ replaced by $({{\mathbf{\widetilde x}}^{(t)}},{\widetilde {\mathbf{L}}^{(t)}})$. Then, the probability of finding at least one approximate solution $({{\mathbf{\widetilde x}}^{(t)}},{\widetilde {\mathbf{L}}^{(t)}})$ satisfying all the events $\{ \mathcal{E}_k^{(t)}\} _{k = 1}^3$, within $T$ rounding trials, is lower bounded by  
\begin{equation}
\begin{split}
\Pr \left( {\bigcup\limits_{t = 1}^T {\left( {\bigcap\limits_{k = 1}^3 {\mathcal{E}_k^{(t)}} } \right)} } \right) & = 1 - \Pr \left( {\bigcap\limits_{t = 1}^T {\left( {\bigcup\limits_{k = 1}^3 {{{(\mathcal{E}_k^{(t)})}^\mathsf{c}}} } \right)} } \right)   \\
& =  1 - \prod\limits_{t = 1}^T {\Pr \left( {\bigcup\limits_{k = 1}^3 {{{(\mathcal{E}_k^{(t)})}^\mathsf{c}}} } \right)}   \\
& = 1 - {\left[ {\Pr \left( {\bigcup\limits_{k = 1}^3 {\mathcal{E}_k^\mathsf{c}} } \right)} \right]^T}    \\
& = 1 - {\left[ 1 - {\Pr \left( {\bigcap\limits_{k = 1}^3 {\mathcal{E}_k} } \right)} \right]^T}      \\
& \geq 1 - {{\xi '}^T}  ,
\end{split}
\end{equation}
where the first and fourth equalities are because of De Morgan's law, the second and third equalities follows from the i.i.d. assumption of rounding trials, while the last inequality is due to \eqref{equation:overall_guarantee_constraints}. It is not hard to see that, given $\xi ' \in (0,3/4]$, this probability tends to $1$ as $T \to \infty $.\footnote{In practice, if Algorithm \ref{algorithm:LPR-RA} fails to find a feasible solution despite performing a relatively large number of rounding trials, then we can use Algorithm \ref{algorithm:LPR-GA} that is guaranteed to find a feasible solution.}
\end{remark}
\vspace{2mm}

Concerning the complexity of Algorithm \ref{algorithm:LPR-RA}, the LPR problem requires $O({N^{3.5}})$ time, while the computation of $({\mathbf{\widetilde x}},\widetilde {\mathbf{L}})$ takes $\Theta (N)$ time. As a result, the total complexity of LPR-RA is $O({N^{3.5}} + N) = O({N^{3.5}})$, that is, asymptotically the same (up to a constant) as that of LPR and LPR-GA. In case of $T$ i.i.d. rounding trials (see Remark \ref{remark:Performance_improvement}), the computational complexity increases to $O({N^{3.5}} + NT)$. Finally, the complexity and performance of all optimization algorithms are summarized in Table \ref{table:Complexity_Performance}.\footnote{The convergence of Algorithms \ref{algorithm:LPR-GA} and \ref{algorithm:LPR-RA} is theoretically guaranteed, since the number of for/while-loop iterations is finite for any given $N$ (recall that $M \leq N$).}

\begin{table}[!t]
\caption{Complexity \& Performance of Optimization Algorithms} 
\centering
\renewcommand{\arraystretch}{1.8}
\begin{threeparttable}
\begin{tabular}{|c|c|c|}
\hline
\makecell{\textbf{Optimization} \\ \textbf{algorithm}} & \makecell{\textbf{Computational} \\ \textbf{complexity}} & \makecell{\textbf{Performance} \\ \textbf{guarantee}} \\ \hline
\makecell{Exhaustive \\ Enumeration} & \makecell{$\Theta \left( \sum\limits_{\begin{subarray}{l} \mathcal{I} \subseteq \mathcal{N} \\ \left| \mathcal{I} \right| \leq M \end{subarray}}  {\left| \mathcal{I} \right| \prod\limits_{i \in \mathcal{I}} {R_i} } \right)$, \\ where, for all $n \in \mathcal{N}$, \\ $R_n=L_n^{\max} - L_n^{\min} + 1$ } & \makecell{Globally optimal \\ solution}  \\ \hline
LPR & $O({N^{3.5}})$ & Lower bound  \\ \hline
\makecell{LPR-GA \\ (Algorithm \ref{algorithm:LPR-GA})} & $O({N^{3.5}})$ & Feasible solution  \\ \hline
\makecell{LPR-RA \\ (Algorithm \ref{algorithm:LPR-RA})}  & \makecell{$O({N^{3.5}})$ or \\ $O({N^{3.5}} + NT)$, \\ where $T$ is the number \\ of i.i.d. rounding trials} & \makecell{Approximate solution \\ with probabilistic \\ performance guarantees \\ (see Theorems \ref{theorem:Expectation_guarantees} and \ref{theorem:Deviation_guarantees}  \\ as well as Remark \ref{remark:Performance_improvement})} \\ \hline
\makecell{AEGA\tnote{*} \\ (Benchmark I)} & $O(N\log N)$ & Feasible solution \\ \hline
\makecell{MEGA\tnote{*} \\ (Benchmark II)}  & $O(N\log N)$ & Feasible solution \\ \hline
\end{tabular}
\begin{tablenotes}
\item[*] These heuristic algorithms are described in Section \ref{subsection:benchmarks}.
\end{tablenotes}
\end{threeparttable}
\label{table:Complexity_Performance}
\end{table}

\section{Numerical Results and Discussion} \label{section:Numerical_results}

The main objective of this section is twofold: the first is to compare the performance of optimization algorithms (in terms of the upper bound of system outage probability) and the second is the comparison between FD and HD schemes. 

\subsection{Simulation Setup}

\begin{table}[!t]
\caption{Simulation Parameters} 
\centering
\renewcommand{\arraystretch}{1.5}
\resizebox{0.495\textwidth}{!}{
\begin{tabular}{|c|c|}
\hline
\textbf{Parameter}  & \textbf{Value}  \\ \hline
Positions of UEs  & $(0,0)$, $(100,0)$ \\ \hline
Transmit power of UEs, $P$ & $25\ \text{dBm}$ \\ \hline
Channel variance, $\sigma^2$ & $1$ \\ \hline
Residual-LI power, $\sigma_{\text{LI}}^2$ & $-70\ \text{dBm}$ \\ \hline
Noise power, $\sigma_w^2$ & $-80\ \text{dBm}$ \\ \hline
SINR threshold, $\gamma_\text{th}$ & $8\ \text{dB}$ \\ \hline
\makecell{Path-loss model parameters (where \\ the distance is measured in meters)} & ${A_0} = 1$, $\alpha  = 2.7$  \\ \hline
IRS location & \makecell{${\operatorname{Uniform}}\left( {[30,70] \times [20,40]} \right)$ or \\ ${\operatorname{Uniform}}\left( {[30,70] \times [ - 40, - 20]} \right)$ \\ with probability $1/2$ } \\ \hline
\makecell{Number of available \\ IRS locations, $N$} & $25$ \\ \hline
\makecell{Maximum number of \\ installed IRSs, $M$} & $7$ \\ \hline
\makecell{Minimum number of reflecting \\ elements for each IRS, \\ $L_n^{\min} = {L^{\min}}$, $\forall n \in \mathcal{N}$} & $5$ \\ \hline
\makecell{Maximum number of reflecting \\ elements for each IRS, \\ $L_n^{\max} = {L^{\max}}$, $\forall n \in \mathcal{N}$} & $40$ \\ \hline
\makecell{Maximum total number of \\ reflecting elements, $L_{{\text{tot}}}^{\max}$} & $250$ \\ \hline
\makecell{Maximum total IRS \\ installation cost, $C_{{\text{tot}}}^{\max}$} & $75$ \\ \hline
IRS fixed deployment cost, ${c_n}$ & ${\operatorname{Uniform}}\left( [1,5] \right)$ \\ \hline
IRS cost rate, $\lambda_n$ & ${\operatorname{Uniform}}\left( [0.1,0.5] \right)$ \\ \hline
\makecell{Maximum number of i.i.d. rounding \\ trials for LPR-RA, $T_{\max}$}  &  $50$ \\ \hline
\end{tabular} }
\label{table:Simulation_Parameters}
\end{table}

We generate random system configurations, where UE-1 and UE-2 are constantly located at $(0,0)$ and
$(100,0)$, respectively, while each IRS location is uniformly distributed either inside the rectangle $[30,70] \times [20,40]$ or $[30,70] \times [-40,-20]$, with probability $1/2$ of being in each rectangle.\footnote{Here, the IRSs are distributed in the 2D-space for the sake of simplicity. Nevertheless, the proposed methodology is straightforwardly applicable to 3D-space scenarios, where the IRSs may be located at different heights.} All channel coefficients are generated according to Remark \ref{remark:Rayleigh_fading} (i.e., Rayleigh fading). Unless otherwise stated, the system parameters are presented in Table \ref{table:Simulation_Parameters}; the positions of UEs and IRSs are on the x-y plane and all coordinates are expressed in meters. Note that, for a given problem, the IRS locations and $\{c_n,\lambda_n\}_{n \in \mathcal{N}}$ are all fixed (the randomization is only used for problem generation). Since there are no established values of IRS cost parameters yet, the simulations are just indicative to evaluate algorithms' performance. In addition, all figures present average values obtained from $10^3$ independent simulation scenarios. Also, the LPR problem \eqref{equation:LPR_problem} is solved using CVX software \cite{Grant2014} with SDPT3 solver \cite{Toh1999}.

Moreover, for every problem instance, LPR-RA performs a maximum number of i.i.d. rounding trials, $T_{\max}$, in order to increase the probability of achieving a feasible solution. In particular, Algorithm \ref{algorithm:LPR-RA} generates independent approximate solutions successively and terminates either when a feasible solution is found or when the maximum number of trials is reached. In all figure captions, we have included the percentage of problems for which LPR-RA returned a feasible solution, within the maximum number of rounding trials.

Despite the fact that a comparison with the optimum value would have been useful, this does not appear in the numerical results (except Fig. \ref{fig:sigma2_noise}) because the complexity of the exhaustive-enumeration algorithm is extremely high. For example, consider a problem with $N = 25$, $M = 7$, $L^{\min} = 5$ and $L^{\max} = 40$ (as shown in Table \ref{table:Simulation_Parameters}). In this case, the exhaustive-enumeration algorithm would require at least $\binom{N}{M} M {R^M} \approx 2.637 \times {10^{17}}$ arithmetic operations, where $R = L^{\max } - L^{\min } + 1$ (cf. Section \ref{subsection:Exhaustive}). If each operation takes approximately $1\ {\mu}\text{s}$, then the algorithm's runtime (for a \emph{single} problem) is roughly $7.325 \times {10^7}\ \text{hr} \approx 3.052 \times {10^6}\ \text{d} \approx 8.362 \times {10^3}\ \text{yr}$, which is obviously prohibitive. 

Nevertheless, we definitely know that the minimum value always lies between LPR (lower bound) and LPR-GA (upper bound, since its solution is guaranteed to be feasible), that is, ${G^\dag} \leq {G^ \star} \leq G'$. Similar inequalities also hold for the upper bound of system outage probability, ${\overline P_{{\text{out}}}}( \cdot )$, due to the monotonicity of exponential function; recall that $G( \cdot ) = \log \left( {{{\overline P}_{{\text{out}}}}( \cdot )} \right)$.

\subsection{Baseline Schemes (Benchmarks)} \label{subsection:benchmarks}

In order to evaluate the performance of the proposed algorithms, we consider (besides the lower bound obtained from LPR) two baseline schemes: 
\begin{itemize}
\item \textit{Average-element greedy algorithm (AEGA):} First, the number of reflecting elements ${L_n}$ is set equal to $L_n^{{\text{avg}}} \triangleq \left\lceil {\tfrac{1}{2}(L_n^{\min } + L_n^{\max })} \right\rceil  \in \{ L_n^{\min }, \ldots ,L_n^{\max }\} $ for all $n \in \mathcal{N}$. Then, we sort the IRS locations in ascending order in terms of the product ${\beta _n}L_n^{{\text{avg}}}$ $(\leq 0)$. Let $({\sigma _1}, \ldots ,{\sigma _N}) \in {\Sigma _\mathcal{N}}$ be the order of IRS locations after sorting, where ${\Sigma _\mathcal{N}}$ is the set of all permutations of $\mathcal{N}$, therefore ${\beta _{{\sigma _1}}}L_{{\sigma _1}}^{{\text{avg}}} \leq  \cdots  \leq {\beta _{{\sigma _N}}}L_{{\sigma _N}}^{{\text{avg}}}$. Finally, AEGA follows the steps 12--20 of Algorithm \ref{algorithm:LPR-GA} (LPR-GA) to compute the binary vector $\mathbf{x}$. In essence, this algorithm can be seen as a greedy procedure that consecutively selects the IRS location inducing the maximum decrease in the objective value (i.e., the upper bound of system outage probability), among the IRS locations not selected yet, while satisfying all constraints. 

\item \textit{Maximum-element greedy algorithm (MEGA):} This algorithm follows the same procedure as AEGA by replacing $L_n^{{\text{avg}}}$ with $L_n^{\max }$, for all $n \in \mathcal{N}$.

\end{itemize}

Both methods are heuristic algorithms (i.e., without a priori performance guarantees) with computational complexity $O(N\log N)$ because of the sorting procedure. In addition, they always find a feasible solution due to their design. 

\subsection{Performance Comparison of Optimization Algorithms}

Subsequently, the performance of optimization algorithms is examined by varying the maximum total IRS installation cost, the maximum number of reflecting elements, the number of available IRS locations, the SINR threshold as well as the noise power.

\subsubsection{Impact of the Maximum Total IRS Installation Cost}

\begin{figure}[!t]
\centering
\includegraphics[width=3.5in]{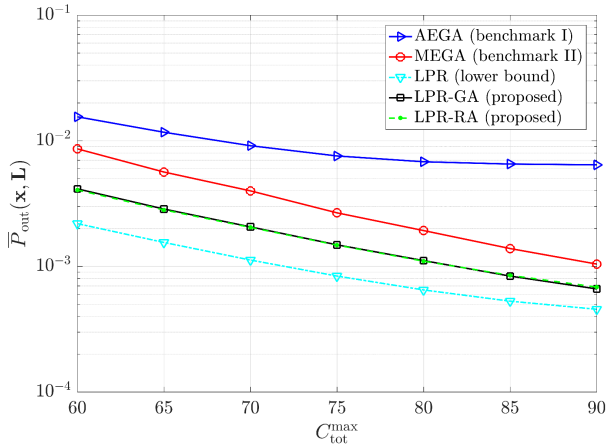} 
\caption{Upper bound of system outage probability versus the maximum total IRS installation cost. The percentage of problems for which LPR-RA has achieved a feasible solution is $[98.5,\ 98.4,\ 99.3,\ 98.6,\ 99.1,\ 99.2,\ 99.5]\%$ for each value of $C_{\text{tot}}^{\max}$, respectively.}
\label{fig:C_tot_max}
\end{figure}

Fig. \ref{fig:C_tot_max} shows the upper bound of system outage probability, against the maximum total IRS installation cost, achieved by AEGA, MEGA, LPR, LPR-GA (Algorithm \ref{algorithm:LPR-GA}) and LPR-RA (Algorithm \ref{algorithm:LPR-RA}). As expected, the outage probability is a nonincreasing function of $C_{{\text{tot}}}^{\max }$ for all algorithms, since larger $C_{{\text{tot}}}^{\max }$ translates to a less restricted feasible set. Furthermore, the proposed algorithms, LPR-GA and LPR-RA, achieve higher performance than the two benchmarks, AEGA and MEGA (with AEGA having the worst performance). It is interesting to observe that LPR-GA and LPR-RA exhibit very similar (almost identical) performance, which is close to the lower bound of LPR. Intuitively, we anticipate such behavior because both algorithms rely on LPR. Finally, LPR-RA demonstrates extremely high percentage (above $98\%$) of achieving a feasible solution within ${T_{\max }} = 50$ rounding trials.

\begin{figure}[!t]
\centering
\includegraphics[width=3.5in]{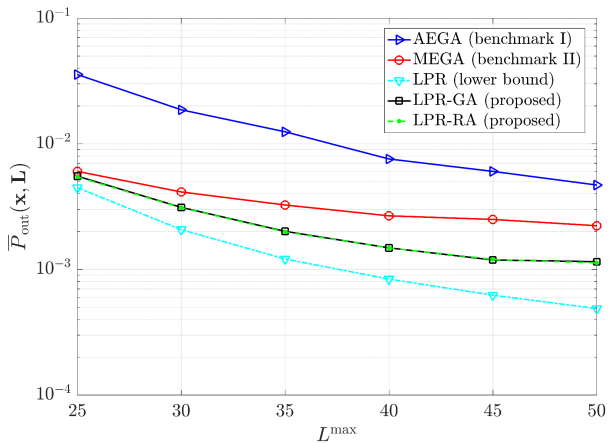} 
\caption{Upper bound of system outage probability versus the maximum number of reflecting elements. The percentage of problems for which LPR-RA has found a feasible solution is $[98.7,\ 98.4,\ 98.1,\ 98.3,\ 99.6,\ 98.3]\%$ for each value of $L^{\max}$, respectively.}
\label{fig:L_max}
\end{figure}

\begin{figure}[!t]
\centering
\includegraphics[width=3.5in]{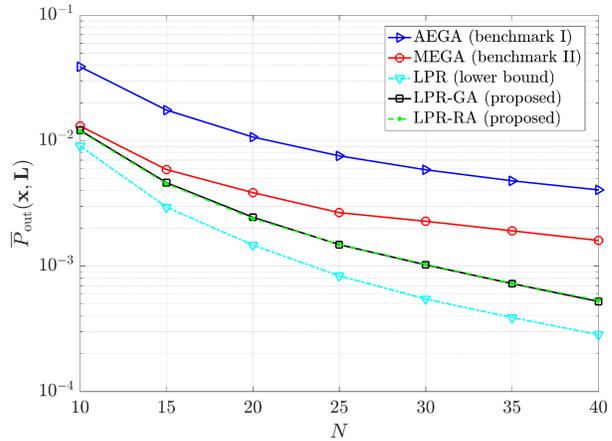} 
\caption{Upper bound of system outage probability versus the number of available IRS locations. The percentage of problems for which LPR-RA has returned a feasible solution is $[97.7,\ 98.2,\ 98.5,\ 98.9,\ 99.1,\ 98.7,\ 99.6]\%$ for each value of $N$, respectively.}
\label{fig:N}
\end{figure}

\subsubsection{Impact of the Maximum Number of Reflecting Elements}

The effect of the maximum number of reflecting elements (for each IRS) on the system outage probability is presented in Fig. \ref{fig:L_max}. Similar conclusions with Fig. \ref{fig:C_tot_max} can be drawn here as well. In addition, MEGA performs slightly worse than LPR-GA and LPR-RA for small values of ${L^{\max }}$ (e.g., 25 and 30), whereas their difference in performance becomes more apparent as ${L^{\max }}$ increases.

\subsubsection{Impact of the Number of Available IRS Locations}

Fig. \ref{fig:N} illustrates the system outage probability as a function of the number of candidate locations for installing an IRS. More specifically, we can observe that, as $N$ increases, all algorithms achieve lower outage probability because there are more options available for the IRS deployment. Once again, the proposed algorithms (LPR-GA and LPR-RA) significantly outperform the baseline schemes, especially when $N$ is relatively large, and remain close to the lower bound (and, thus, to the global minimum) for all values of $N$. The latter demonstrates the robustness of the developed algorithms in terms of the size of the search space (cf. Figs. \ref{fig:C_tot_max} and \ref{fig:L_max}).

\subsubsection{Impact of the SINR Threshold}

\begin{figure}[!t]
\centering
\includegraphics[width=3.5in]{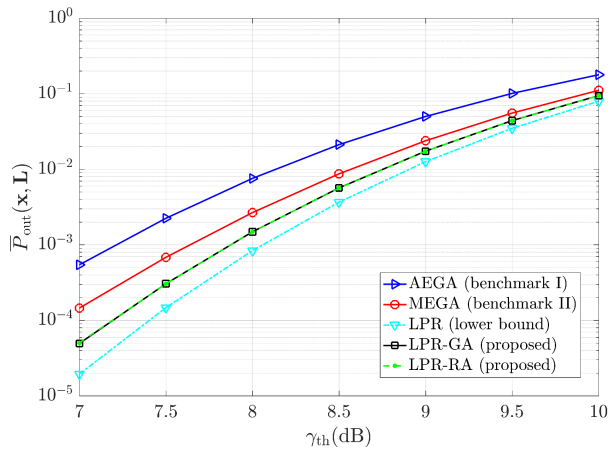} 
\caption{Upper bound of system outage probability versus the SINR threshold. The percentage of problems for which LPR-RA has achieved a feasible solution is $[99.1,\ 98.9,\ 98.6,\ 98.5,\ 99.2,\ 98.4,\ 98.4]\%$ for each value of $\gamma_{\text{th}}$, respectively.}
\label{fig:gamma_th}
\end{figure}

The upper bound of system outage probability versus the SINR threshold is shown in Fig. \ref{fig:gamma_th}. In particular, the outage probability increases with ${\gamma_{{\text{th}}}}$ for all optimization algorithms, which is intuitively expected. Furthermore, it is interesting to note that the distance of the objective value achieved by AEGA, MEGA, LPR-GA, LPR-RA from the lower bound decreases as the SINR threshold increases. Roughly speaking, this means that the achieved solution tends to be globally optimal, especially for LPR-GA and LPR-RA.

\subsubsection{Impact of the Noise Power and Comparison with Exhaustive Search}

\begin{figure}[!t]
\centering
\includegraphics[width=3.5in]{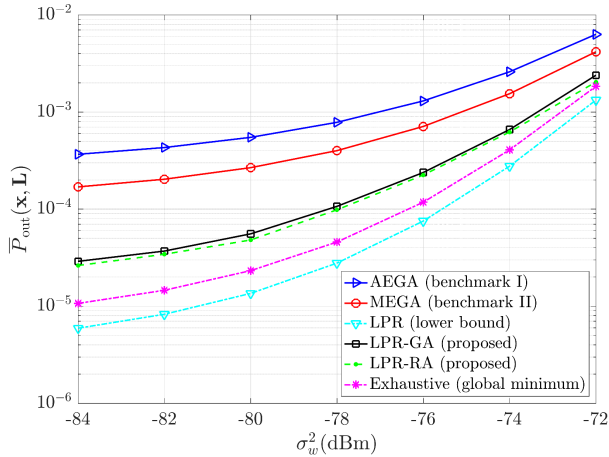} 
\caption{Upper bound of system outage probability versus the noise power at each UE. The percentage of problems for which LPR-RA has returned a feasible solution is $[99,\ 97,\ 98,\ 96,\ 100,\ 98,\ 98]\%$ for each value of $\sigma_w^2$, respectively.}
\label{fig:sigma2_noise}
\end{figure}

Here, we also consider the exhaustive-enumeration algorithm under small-scale setups with the following parameters: $\gamma_\text{th} = 3\ \text{dB}$, $N = 7$, $M = 4$, $L^{\min} = 35$, $L^{\max} = 50$, $L_\text{tot}^{\max} = 115$, $C_\text{tot}^{\max} = 30$. In Fig. \ref{fig:sigma2_noise}, it is clear that the upper bound of outage probability increases with the noise power, for all algorithms. In spite of the fact that the proposed algorithms do not achieve the global minimum (recall that problem \eqref{equation:discrete_problem} is NP-hard, according to Theorem \ref{theorem:NP_hardness}), their performance is again much higher than AEGA and MEGA. Moreover, the average number of selected IRS locations, the average total number of reflecting elements, and the average total IRS installation cost are respectively: $[1.4, 60.6, 21.2]$ for AEGA, $[1.2, 61.5, 20.8]$ for MEGA, $[1.7, 86, 21.9]$ for LPR-GA, $[1.7, 85.9, 22]$ for LPR-RA, and $[2.1, 92.7, 24.5]$ for exhaustive enumeration. We can observe that LPR-GA/RA selects more IRS locations and reflecting elements at the expense of higher cost compared to AEGA/MEGA; this is also true if we compare exhaustive enumeration with LPR-GA/RA.

\subsection{Comparison between FD and HD Systems}

Afterwards, the optimization algorithms are used in order to make meaningful comparisons between FD and HD wireless technologies.\footnote{For details on the HD scheme, see Section \ref{subsection:HD}.}

\subsubsection{Impact of the Transmit Power of UEs}

\begin{figure}[!t]
\centering
\includegraphics[width=3.5in]{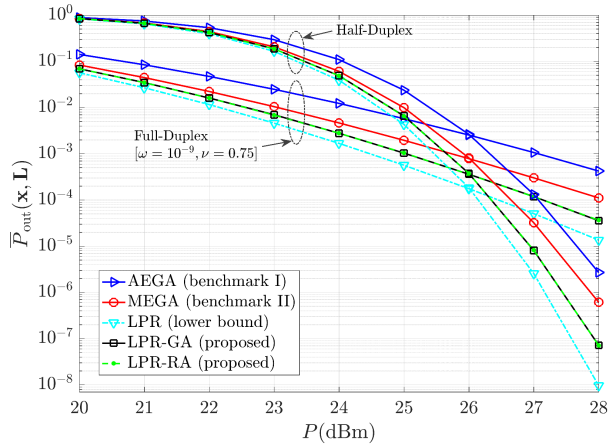} 
\caption{Upper bound of system outage probability versus the transmit power of UEs for $\gamma _{\text{th}} = 9\ \text{dB}$; $\sigma _{\text{LI}}^2 = \omega {P^\nu}$ in FD scheme, whereas $\sigma _{\text{LI}}^2 = 0$ in HD scheme. The percentage of problems for which LPR-RA has found a feasible solution is $[98.3,\ 98.6,\ 98.6,\ 98.0,\ 99.1,\ 99.1,\ 98.6,\ 98.8,\ 99.2]\%$ in FD system and $[97.7,\ 98.3,\ 98.3,\ 99.1,\ 99.1,\ 98.3,\ 98.7,\ 99.0,\ 99.4]\%$ in HD system for each value of $P$, respectively.}
\label{fig:P}
\end{figure}

Firstly, let us examine the effect of the transmit power of UEs on the system outage probability. According to Fig. \ref{fig:P}, the proposed optimization algorithms perform much better than the two benchmarks, especially for high transmit power, in both HD and FD schemes. In addition, we can observe that FD is superior to HD system (in terms of outage probability) when $P < 26\ \text{dBm}$, irrespective of the algorithm used for comparison. On the other hand, FD is inferior to HD scheme when $P > 26\ \text{dBm}$.

\subsubsection{Impact of the Residual-LI Power}

\begin{figure}[!t]
\centering
\includegraphics[width=3.5in]{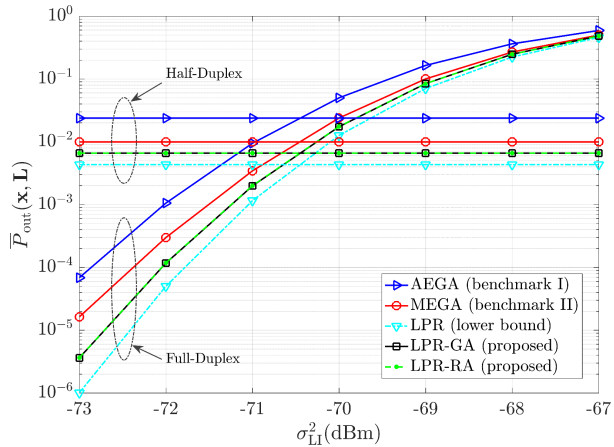} 
\caption{Upper bound of system outage probability versus the residual-LI power at each UE for $\gamma_{\text{th}} = 9\ \text{dB}$; the transmit power is fixed at $P = 25\ \text{dBm}$. The percentage of problems for which LPR-RA has returned a feasible solution is $[99.0,\ 98.9,\ 98.7,\ 98.9,\ 98.7,\ 98.5,\ 98.9]\%$ in FD system for each value of $\sigma_{\text{LI}}^2$, respectively, and $98.6\%$ in HD system regardless of $\sigma_{\text{LI}}^2$.}
\label{fig:sigma2_LI}
\end{figure}

Secondly, we study how the residual-LI power affects the system outage probability, assuming constant transmit power. Based on Fig. \ref{fig:sigma2_LI}, it is obvious that LPR-GA and LPR-RA again show better performance compared to the baseline schemes, not only in FD but also in HD scenario. Moreover, for the FD scheme, the upper bound of system outage probability increases rapidly with the residual-LI power. Finally, FD outperforms HD system when $\sigma_{{\text{LI}}}^2$ is approximately less than $-70.4\ \text{dBm}$, whereas HD is preferable when $\sigma_{{\text{LI}}}^2$ is greater than $-70.4\ \text{dBm}$ (regardless of the comparison algorithm). In other words, FD is more beneficial than HD technology, provided that the LI at each UE is sufficiently suppressed.

\section{Conclusion and Future Work} \label{section:Conclusion}

In this paper, we have dealt with the minimization of outage probability in a two-way FD communication system assisted by multiple IRSs. In particular, we have transformed the joint IRS location and size optimization problem into a discrete problem, which turned out to be NP-hard. In order to overcome this difficulty, two efficient (polynomial-time) algorithms have been proposed, which are based on the solution of LPR. The first is a deterministic greedy method, while the second is a randomized approximation algorithm. According to the numerical results, the developed algorithms have shown much higher performance than the benchmarks. Their achieved objective values have also been close to the lower bound and, therefore, to the global minimum. Moreover, we have observed that FD outperforms HD scheme, provided that the LI at both UEs is adequately mitigated; this is in line with our intuition. 

Finally, we mention several challenging research directions: i) extension to multiple-antenna systems with multiple users, ii) analysis of IRS-activation schemes with lower signaling overhead using, for example, partial CSI knowledge and iii) investigation of IRS phase-adjustment errors due to imperfect CSI or phase-quantization errors (finite phase-shift resolution).

\appendices

\section{Proof of Theorem \ref{theorem:NP_hardness}} \label{appendix:NP_hardness}

It is sufficient to show that there is a special case of problem \eqref{equation:discrete_problem} which is NP-hard. Let us consider the following case: $L_n^{\min } = L_n^{\max } = L_n^{\circ} > 0$ (hence, ${L_n} = L_n^{ \circ }$) for all $n \in \mathcal{N}$, $M = N$ (so, the IRS-cardinality constraint can be omitted, because $\sum\nolimits_{n \in \mathcal{N}} {{x_n}}  \leq N$ for all ${\mathbf{x}} \in {\{ 0,1\} ^N}$) and $L_{{\text{tot}}}^{\max } = \sum\nolimits_{n \in \mathcal{N}} {L_n^{\max }} $ (therefore, the constraint of maximum total number of reflecting elements can be omitted, since $\sum\nolimits_{n \in \mathcal{N}} {{x_n}{L_n}}  \leq \sum\nolimits_{n \in \mathcal{N}} {{L_n}}  \leq \sum\nolimits_{n \in \mathcal{N}} {L_n^{\max }} $ for all ${\mathbf{x}} \in {\{ 0,1\} ^N}$ and ${\mathbf{L}} \in { \times _{n \in \mathcal{N}}}\{ L_n^{\min }, \ldots ,L_n^{\max }\}$). As a consequence, problem \eqref{equation:discrete_problem} reduces to   
\begin{subequations} 
\begin{alignat}{3}
  & \mathop {\min }\limits_{{{\mathbf{x}}}} & \quad & \sum\limits_{n \in \mathcal{N}} {({\beta _n}L_n^{\circ }){x_n}}     \\
  & ~\text{s.t.} & & {x_n} \in \{ 0,1\} ,\;\;\forall n \in \mathcal{N} \\
  & & & \sum\limits_{n \in \mathcal{N}} {({c_n} + {\lambda _n}L_n^{\circ }){x_n}} \leq C_{{\text{tot}}}^{\max } .
\end{alignat}
\end{subequations}

By converting it into a maximization problem, we obtain 
\begin{subequations} \label{equation:special_discrete_problem}
\begin{alignat}{3}
  & \mathop {\max }\limits_{{{\mathbf{x}}}} & \quad & \sum\limits_{n \in \mathcal{N}} {{\varphi _n}{x_n}}    \\
  & ~\text{s.t.} & & {x_n} \in \{ 0,1\} ,\;\;\forall n \in \mathcal{N} \\
  & & & \sum\limits_{n \in \mathcal{N}} {{\theta _n}{x_n}}  \leq C_{{\text{tot}}}^{\max }  ,
\end{alignat}
\end{subequations}
where ${\varphi _n} =  - {\beta _n}L_n^{\circ } \geq 0$ and ${\theta _n} = {c_n} + {\lambda _n}L_n^{\circ }$ for every $n \in \mathcal{N}$. A crucial observation here is that each coefficient in ${\{ {\varphi _n},{\theta _n}\} _{n \in \mathcal{N}}}$ can be \emph{any} positive integer. To prove this, let ${\kappa _n},{\mu _n} > 0$ be arbitrary integers. Then, we can find a problem instance such that ${\rho _n} = {\gamma _{{\text{th}}}} \big/ {\left( {{F_n^{ - 1}}({e^{ - {\kappa _n}/L_n^{\circ }}})} \right)^2}$, $0 \leq {c_n} \leq {\mu _n}$ and ${\lambda _n} = ({\mu _n} - {c_n})/L_n^{\circ } \geq 0$ for all $n \in \mathcal{N}$, where ${F_n^{ - 1}}( \cdot )$ is the inverse function of $F_n( \cdot )$ given by \eqref{equation:CDF_single_element}; the existence of ${F_n^{ - 1}}( \cdot )$ is guaranteed, because $F_n( \cdot )$ is continuous and (strictly) increasing. As a result, ${\beta _n} = \log \left( {F_n(\sqrt {{\gamma _{{\text{th}}}}/{\rho _n}} )} \right) =  - {\kappa _n}/L_n^{\circ } < 0$, thus ${\varphi _n} = {\kappa _n}$, and ${\theta _n} = {\mu _n}$ as well. 

Furthermore, if ${\{ {\varphi _n},{\theta _n}\} _{n \in \mathcal{N}}}$ and $C_{{\text{tot}}}^{\max }$ are all restricted to be positive integers, then problem \eqref{equation:special_discrete_problem} becomes identical to the \emph{knapsack problem} which is known to be NP-hard \cite{Papadimitriou}; this completes the proof.

\section{Proof of Theorem \ref{theorem:Expectation_guarantees}} \label{appendix:Expectation_guarantees}

First of all, due to \eqref{equation:x_approx}--\eqref{equation:L_approx_b}, we have that 
\begin{equation}
\mathbb{E}({\widetilde x_n}) = 1 \cdot x_n^\dag + 0 \cdot (1 - x_n^\dag)  = x_n^\dag \, ,
\end{equation}
\begin{equation}
\begin{split}
\mathbb{E}\left( {\left. {{{\widetilde L}_n}} \right| x_n^\dag  \ne 0} \right) & = \left( {\left\lfloor {L_n^\dag } \right\rfloor  + 1} \right)\operatorname{frac} (L_n^\dag )  \\
& \;\;\;\;\, + \left\lfloor {L_n^\dag } \right\rfloor \left( {1 - \operatorname{frac} (L_n^\dag )} \right) \\
& =  \operatorname{frac} (L_n^\dag ) + \left\lfloor {L_n^\dag } \right\rfloor  
 = L_n^\dag    \, ,
\end{split}
\end{equation}
\begin{equation}
\begin{split}
\mathbb{E}\left( {\left. {{{\widetilde L}_n}} \right| x_n^\dag  = 0} \right) & = \tfrac{1}{L_n^{\max} - L_n^{\min} + 1} \sum_{i = L_n^{\min}}^{L_n^{\max}} {i}  \\
& = \tfrac{1}{2}(L_n^{\min } + L_n^{\max })   .
\end{split}
\end{equation}

Furthermore, by defining the random variable ${\widetilde z_n} = {\widetilde x_n}{\widetilde L_n}$ for all $n \in \mathcal{N}$ and exploiting the independency of ${\widetilde x_n}$ and ${\widetilde L_n}$, we get the following conditional expectations: 
\begin{equation}
\begin{split}
\mathbb{E}\left( {\left. {{{\widetilde z}_n}} \right|x_n^\dag  \ne 0} \right) & = \mathbb{E}\left( {\left. {{{\widetilde x}_n}} \right| x_n^\dag  \ne 0} \right)\mathbb{E}\left( {\left. {{{\widetilde L}_n}} \right|x_n^\dag  \ne 0} \right) \\
& = x_n^\dag L_n^\dag  = z_n^\dag  \, ,
\end{split}
\end{equation}
\begin{equation}
\begin{split}
\mathbb{E}\left( {\left. {{{\widetilde z}_n}} \right|x_n^\dag  = 0} \right) &= \mathbb{E}\left( {\left. {{{\widetilde x}_n}} \right|x_n^\dag  = 0} \right)\mathbb{E}\left( {\left. {{{\widetilde L}_n}} \right|x_n^\dag  = 0} \right)   \\
& = 0 \cdot \tfrac{1}{2}(L_n^{\min } + L_n^{\max }) = 0 = z_n^\dag  \, ,
\end{split}
\end{equation}
where the latter is based on the fact that: if $x_n^\dag  = 0$, then $z_n^\dag  = 0$ and ${\widetilde x_n} = 0$. As a result, combining the above equations, $\mathbb{E}\left( {{{\widetilde z}_n}} \right) = z_n^\dag $ for all $n \in \mathcal{N}$. 

Because of the linearity of expectation and the feasibility of $({{\mathbf{x}}^\dag },{{\mathbf{z}}^\dag })$, we obtain 
\begin{equation}
\begin{split}
\mathbb{E}\left( {G({\mathbf{\widetilde x}},\widetilde {\mathbf{L}})} \right) & = \mathbb{E}\left( {\sum\limits_{n \in \mathcal{N}} {{\beta _n}{{\widetilde z}_n}} } \right) = \sum\limits_{n \in \mathcal{N}} {{\beta _n}\mathbb{E}({{\widetilde z}_n})}  \\
&  = \sum\limits_{n \in \mathcal{N}} {{\beta _n}z_n^\dag }  = {G^\dag } ,
\end{split}
\end{equation}
\begin{equation}
\mathbb{E}\left( {\sum\limits_{n \in \mathcal{N}} {{{\widetilde x}_n}} } \right) = \sum\limits_{n \in \mathcal{N}} {\mathbb{E}({{\widetilde x}_n})}  = \sum\limits_{n \in \mathcal{N}} {x_n^\dag } \leq M  ,
\end{equation}
\begin{equation}
\begin{split}
\mathbb{E}\left( {{L_{{\text{tot}}}}({\mathbf{\widetilde x}},\widetilde {\mathbf{L}})} \right) & = \mathbb{E}\left( {\sum\limits_{n \in \mathcal{N}} {{{\widetilde z}_n}} } \right) = \sum\limits_{n \in \mathcal{N}} {\mathbb{E}({{\widetilde z}_n})}  \\
& = \sum\limits_{n \in \mathcal{N}} {z_n^\dag }  \leq L_{{\text{tot}}}^{\max }  ,
\end{split}
\end{equation}
\begin{equation}
\begin{split}
\mathbb{E}\left( {{C_{{\text{tot}}}}({\mathbf{\widetilde x}},\widetilde {\mathbf{L}})} \right) & = \mathbb{E}\left( {\sum\limits_{n \in \mathcal{N}} {{c_n}{{\widetilde x}_n}}  + \sum\limits_{n \in \mathcal{N}} {{\lambda _n}{{\widetilde z}_n}} } \right) \\
&  = \sum\limits_{n \in \mathcal{N}} {{c_n}\mathbb{E}({{\widetilde x}_n})}  + \sum\limits_{n \in \mathcal{N}} {{\lambda _n}\mathbb{E}({{\widetilde z}_n})}   \\
&  = \sum\limits_{n \in \mathcal{N}} {{c_n}x_n^\dag }  + \sum\limits_{n \in \mathcal{N}} {{\lambda _n}z_n^\dag }  \leq C_{{\text{tot}}}^{\max }  .
\end{split}
\end{equation}
Therefore, Theorem \ref{theorem:Expectation_guarantees} has been proven.

\section{Proof of Theorem \ref{theorem:Deviation_guarantees}} \label{appendix:Deviation_guarantees}

In order to prove Theorem \ref{theorem:Deviation_guarantees}, we make use of a concentration inequality for sums of independent and bounded random variables, namely, Hoeffding's inequality. This is a generalization of the well-known Chernoff bound, which applies to Bernoulli random variables. 

\vspace{2mm}
\begin{lemma}[Hoeffding's inequality \cite{Hoeffding1963}] \label{lemma:Hoeffding_inequality}
Let ${\{ {X_n}\} _{n \in \mathcal{N}}}$ be a finite set of independent random variables, with ${X_n} \in [{a_n},{b_n}]$ for all $n \in \mathcal{N}$ (where ${a_n} \leq {b_n}$), and $X \triangleq \sum\nolimits_{n \in \mathcal{N}} {{X_n}} $. Then, for any $u > 0$,
\begin{equation}
\Pr \left( {X - \mathbb{E}(X) > u} \right) \leq {e^{ - 2{u^2}/\Delta }}  \, ,
\end{equation}
where $\Delta  = \sum_{n \in \mathcal{N}} {{{({b_n} - {a_n})}^2}}$.  
\end{lemma}
\vspace{2mm}

The probability of approximating the optimum value within a given tolerance ${\epsilon_0} > 0$ is lower bounded by 
\begin{equation}
\resizebox{0.48\textwidth}{!}{$ 
\begin{split}
\Pr \left( {G({\mathbf{\widetilde x}},\widetilde {\mathbf{L}}) \leq {G^ \star } + {\epsilon_0}} \right) & = 1 - \Pr \left( {G({\mathbf{\widetilde x}},\widetilde {\mathbf{L}}) > {G^ \star } + {\epsilon_0}} \right)  \\
& \geq 1 - \Pr \left( {G({\mathbf{\widetilde x}},\widetilde {\mathbf{L}}) > {G^\dag } + {\epsilon_0}} \right)   \\
& \geq 1 - {e^{ - 2{\epsilon_0^2}/{\Delta _0}}}  ,
\end{split} $}
\end{equation}
where ${\Delta _0} = \sum_{n \in \mathcal{N}} {{{({\beta _n}L_n^{\max })}^2}} $. Here, the first inequality is because ${G^\dag } \leq {G^ \star }$, while the second inequality follows from Lemma \ref{lemma:Hoeffding_inequality} by taking advantage of \eqref{equation:expectation_guarantee_objective} and noticing that $G({\mathbf{\widetilde x}},\widetilde {\mathbf{L}}) = \sum_{n \in \mathcal{N}} {{\beta _n}({{\widetilde x}_n}{{\widetilde L}_n})}  = \sum_{n \in \mathcal{N}} {{\beta _n}{{\widetilde z}_n}} $ is the sum of independent random variables ${\{ {\beta _n}{\widetilde z_n}\} _{n \in \mathcal{N}}}$ with ${\beta _n}{\widetilde z_n} \in [{\beta _n}L_n^{\max },0]$; recall that ${\beta _n} \leq 0$. By assuming ${\Delta _0} > 0$ and setting ${\epsilon_0} = \sqrt {{\Delta _0}\log (N + 1)} > 0$, we get \eqref{equation:deviation_guarantees} for $k = 0$.
 
Regarding the probabilistic guarantee for the IRS-cardinality constraint, we have for any ${\epsilon_1} > 0$  
\begin{equation}
\resizebox{0.48\textwidth}{!}{$ 
\begin{split}
\Pr \left( {\sum\limits_{n \in \mathcal{N}} {{{\widetilde x}_n}} \leq M + {\epsilon_1}} \right) & = 1 - \Pr \left( {\sum\limits_{n \in \mathcal{N}} {{{\widetilde x}_n}}  >  M + {\epsilon_1}} \right) \\
& \geq  1 - \Pr \left( {\sum\limits_{n \in \mathcal{N}} {{{\widetilde x}_n}} > \sum\limits_{n \in \mathcal{N}} {x_n^\dag }  + {\epsilon_1}} \right)  \\
& \geq 1 - {e^{ - 2{\epsilon_1^2}/{\Delta _1}}}  ,
\end{split} $}
\end{equation}
where ${\Delta_1} = N$. The two inequalities follows from \eqref{equation:expectation_guarantee_M} and Lemma \ref{lemma:Hoeffding_inequality}. Finally, by choosing ${\epsilon_1} = \sqrt {{\Delta _1}\log (N + 1)} > 0$ we obtain \eqref{equation:deviation_guarantees} for $k = 1$. 

Moreover, based on \eqref{equation:expectation_guarantee_L_tot_max} and Lemma \ref{lemma:Hoeffding_inequality}, we obtain for any ${\epsilon_2} > 0$ 
\begin{equation} 
\resizebox{0.48\textwidth}{!}{$
\begin{split}
\Pr \left( {{L_{{\text{tot}}}}({\mathbf{\widetilde x}},\widetilde {\mathbf{L}}) \leq L_{{\text{tot}}}^{\max } + {\epsilon_2}} \right) & = 1 - \Pr \left( {{L_{{\text{tot}}}}({\mathbf{\widetilde x}},\widetilde {\mathbf{L}}) > L_{{\text{tot}}}^{\max } + {\epsilon_2}} \right)  \\
& \geq  1 - \Pr \left( {{L_{{\text{tot}}}}({\mathbf{\widetilde x}},\widetilde {\mathbf{L}}) > \sum\limits_{n \in \mathcal{N}} {z_n^\dag }  + {\epsilon_2}} \right)  \\
& \geq  1 - {e^{ - 2{\epsilon_2^2}/{\Delta _2}}}   ,
\end{split} $}
\end{equation}
where ${\Delta _2} = \sum_{n \in \mathcal{N}} {{{(L_n^{\max })}^2}} $. Note that ${L_{{\text{tot}}}}({\mathbf{\widetilde x}},\widetilde {\mathbf{L}}) = \sum_{n \in \mathcal{N}} {{{\widetilde x}_n}{{\widetilde L}_n}}  = \sum_{n \in \mathcal{N}} {{{\widetilde z}_n}} $ is the sum of independent random variables ${\{ {\widetilde z_n}\} _{n \in \mathcal{N}}}$ with ${\widetilde z_n} \in [0,L_n^{\max }]$. Inequality \eqref{equation:deviation_guarantees} for $k = 2$ can be easily derived, assuming ${\Delta _2} > 0$ and setting ${\epsilon_2} = \sqrt {{\Delta _2}\log (N + 1)} > 0$. 

Although ${\widetilde x_n}$ and ${\widetilde z_n} = {\widetilde x_n}{\widetilde L_n}$ are not independent random variables, ${C_{{\text{tot}}}}({\mathbf{\widetilde x}},\widetilde {\mathbf{L}}) = \sum_{n \in \mathcal{N}} {{c_n}{{\widetilde x}_n}}  + \sum_{n \in \mathcal{N}} {{\lambda _n}({{\widetilde x}_n}{{\widetilde L}_n})}$ can be written as the sum of independent random variables ${\{ ({c_n} + {\lambda _n}{\widetilde L_n}){\widetilde x_n}\} _{n \in \mathcal{N}}}$, i.e., ${C_{{\text{tot}}}}({\mathbf{\widetilde x}},\widetilde {\mathbf{L}}) = \sum_{n \in \mathcal{N}} {({c_n} + {\lambda _n}{{\widetilde L}_n}){{\widetilde x}_n}}$, with $({c_n} + {\lambda _n}{\widetilde L_n}){\widetilde x_n} \in [0,{c_n} + {\lambda _n}L_n^{\max }]$. As a consequence, for any ${\epsilon_3} > 0$, the probability of the event ${\mathcal{E}_3}$ can be lower bounded by 
\begin{equation} 
\resizebox{0.48\textwidth}{!}{$ 
\begin{split}
\Pr \left( {{C_{{\text{tot}}}}({\mathbf{\widetilde x}},\widetilde {\mathbf{L}}) \leq C_{{\text{tot}}}^{\max } + {\epsilon_3}} \right) & = 1 - \Pr \left( {{C_{{\text{tot}}}}({\mathbf{\widetilde x}},\widetilde {\mathbf{L}}) > C_{{\text{tot}}}^{\max } + {\epsilon_3}} \right)  \\
& \geq 1 - \Pr \left( { {{C_{{\text{tot}}}}({\mathbf{\widetilde x}},\widetilde {\mathbf{L}}) > \sum\limits_{n \in \mathcal{N}} {{c_n}x_n^\dag } } } \right. \\
& \quad  \left. { + \sum\limits_{n \in \mathcal{N}} {{\lambda _n}z_n^\dag }  + {\epsilon_3} } \right)  \\
& \geq  1 - {e^{ - 2{\epsilon_3^2}/{\Delta _3}}}  ,
\end{split} $}
\end{equation}
where ${\Delta _3} = \sum_{n \in \mathcal{N}} {{{({c_n} + {\lambda _n}L_n^{\max })}^2}}$, while the two inequalities result from the combination of \eqref{equation:expectation_guarantee_C_tot_max} with Lemma \ref{lemma:Hoeffding_inequality}. Assuming ${\Delta _3} > 0$ and choosing ${\epsilon_3} = \sqrt {{\Delta _3}\log (N + 1)} > 0$, we have \eqref{equation:deviation_guarantees} for $k = 3$. Hence, \eqref{equation:deviation_guarantees} has been proven for all $k \in \{ 0,1,2,3\}$. 

Now, let us consider inequality \eqref{equation:overall_guarantee_constraints}. By applying \emph{De Morgan's law} and the \emph{union bound theorem}, we obtain  
\begin{equation}
\Pr \left( {\bigcap\limits_{k = 1}^3 {{\mathcal{E}_k}} } \right) = 1 - \Pr \left( {{{\left( {\bigcap\limits_{k = 1}^3 {{\mathcal{E}_k}} } \right)}^{\mathsf{c}}}} \right) = 1 - \Pr \left( {\bigcup\limits_{k = 1}^3 {\mathcal{E}_k^\mathsf{c}} } \right) ,
\end{equation}
\begin{equation}
\Pr \left( {\bigcup\limits_{k = 1}^3 {\mathcal{E}_k^\mathsf{c}} } \right) \leq \sum\limits_{k = 1}^3 {\Pr (\mathcal{E}_k^\mathsf{c})}  = \sum\limits_{k = 1}^3 {(1 - \Pr ({\mathcal{E}_k}))}  .
\end{equation}
Subsequently, by combining the above equations and leveraging \eqref{equation:deviation_guarantees}, we deduce 
\begin{equation}
\begin{split}
\Pr \left( {\bigcap\limits_{k = 1}^3 {{\mathcal{E}_k}} } \right) & \geq 1 - \sum\limits_{k = 1}^3 {(1 - \Pr ({\mathcal{E}_k}))}  \\
& \geq 1 - {3 \xi} = 1 - {\xi'}  .
\end{split}
\end{equation}

Finally, inequality \eqref{equation:overall_guarantee_objective_constraints} can be derived by following similar steps as above, so Theorem \ref{theorem:Deviation_guarantees} follows immediately.


\end{document}